\documentclass[bibnotes,amsmath,amssymb,showpacs,floatfix,superscriptaddress,reprint]{revtex4-1}
\usepackage{amsmath,empheq}
\usepackage{amsthm}
\usepackage{amsfonts}
\usepackage{amssymb}
\usepackage{amsxtra}
\usepackage{mathtools}
\usepackage{xcolor}
\usepackage{subfigure}
\usepackage{dcolumn}
\usepackage{mathrsfs}
\usepackage{tikz}
\usepackage{float}
\usepackage{bm}
\usepackage[breaklinks=true,colorlinks,citecolor=blue,linkcolor=blue,urlcolor=blue]{hyperref}

\DeclareMathAlphabet{\bi}{OML}{cmm}{b}{it}
\def\be{\begin{equation}}
	\def\ee{\end{equation}}
\def\bearr{\begin{eqnarray}}
	\def\eearr{\end{eqnarray}}

\begin{document}
	\title{Coherent optical control of quantum Hall edge states}
	\author{Ashutosh Singh}
	\email{asingh.n19@gmail.com}
	\affiliation{Department of Physics and Astronomy, Texas A\&M University, College Station, TX, 77843 USA}
 \author{Maria Sebastian}
\affiliation{Department of Physics and Astronomy, Texas A\&M University, College Station, TX, 77843 USA}
\author{Mikhail Tokman}
\affiliation{Department of Electrical and Electronic Engineering and Schlesinger Knowledge Center for Compact Accelerators
and Radiation Sources, Ariel University, 40700 Ariel, Israel}

\author{Alexey Belyanin}%
\email{belyanin@tamu.edu}
\affiliation{Department of Physics and Astronomy, Texas A\&M University, College Station, TX, 77843 USA}

	\date{\today}
	
	\begin{abstract}
		Current carrying chiral edge states in quantum Hall systems have fascinating properties that are usually studied by electron spectroscopy and interferometry. Here we demonstrate that electron occupation, current, and electron coherence in chiral edge states can be selectively probed and controlled by low-energy electromagnetic radiation in the microwave to infrared range without affecting electron states in the bulk or destroying quantum Hall effect conditions in the bulk of the sample. Both linear and nonlinear optical control is possible due to inevitable violation of adiabaticity and inversion symmetry breaking for electron states near the edge.  This opens up new pathways for frequency- and polarization-selective spectroscopy and control of individual edge states.

	\end{abstract}
	
	%
	\maketitle
	%
\section{Introduction}

The Quantum Hall (QH) effect is one of the most studied phenomena in condensed matter physics \cite{QHE}, with far-reaching applications in many other areas. One of the most important insights in the QH effect physics was the intricate relation between the electron states in the insulating bulk of the sample and the current carrying chiral edge states  \cite{Halperin_Edge, MacDonald_edge} which have been extensively studied with real space imaging and momentum resolved electron spectroscopy \cite{PhysRevX.4.011014,Marguerite2019, PhysRevB.107.115426,Li2013, Patlatiuk2018, Kim2021}. Moreover, coherence of unidirectional electron transport in QH edge states stimulated massive research in QH edge state interferometry. Various types of electron interferometers have been implemented, both in conventional semiconductor quantum wells and in graphene samples, and for both integer and fractional statistics of carriers \cite{PhysRevLett.62.2523, Chamon, Ji2003, PhysRevB.73.245311, PhysRevLett.96.016802,PhysRevLett.97.186803, PhysRevB.79.241304,PhysRevB.82.085321,Corentin2021, PhysRevB.105.165310, Nakamura2020, Ron2021, nak23}. 

Terahertz optical spectroscopy of Landau-quantized electron states in two-dimensional (2D) electron gas is of course yet another massive field of research. However, its utility in probing or manipulating the electron states under the conditions of QH effect is highly problematic, because resonant optical transitions between bulk Landau levels will lead to nonequilibrium carrier population, which therefore enables nonzero bulk DC conductivity across the sample. As was argued in Ref.~\cite{doi:10.1126/science.abl5818}, even vacuum cavity fields under the ultra-strong coupling conditions could break the topological protection of the integer QH effect and destroy some of the high quantum number plateaux. 

\begin{figure}[!]
\includegraphics[width = \linewidth]{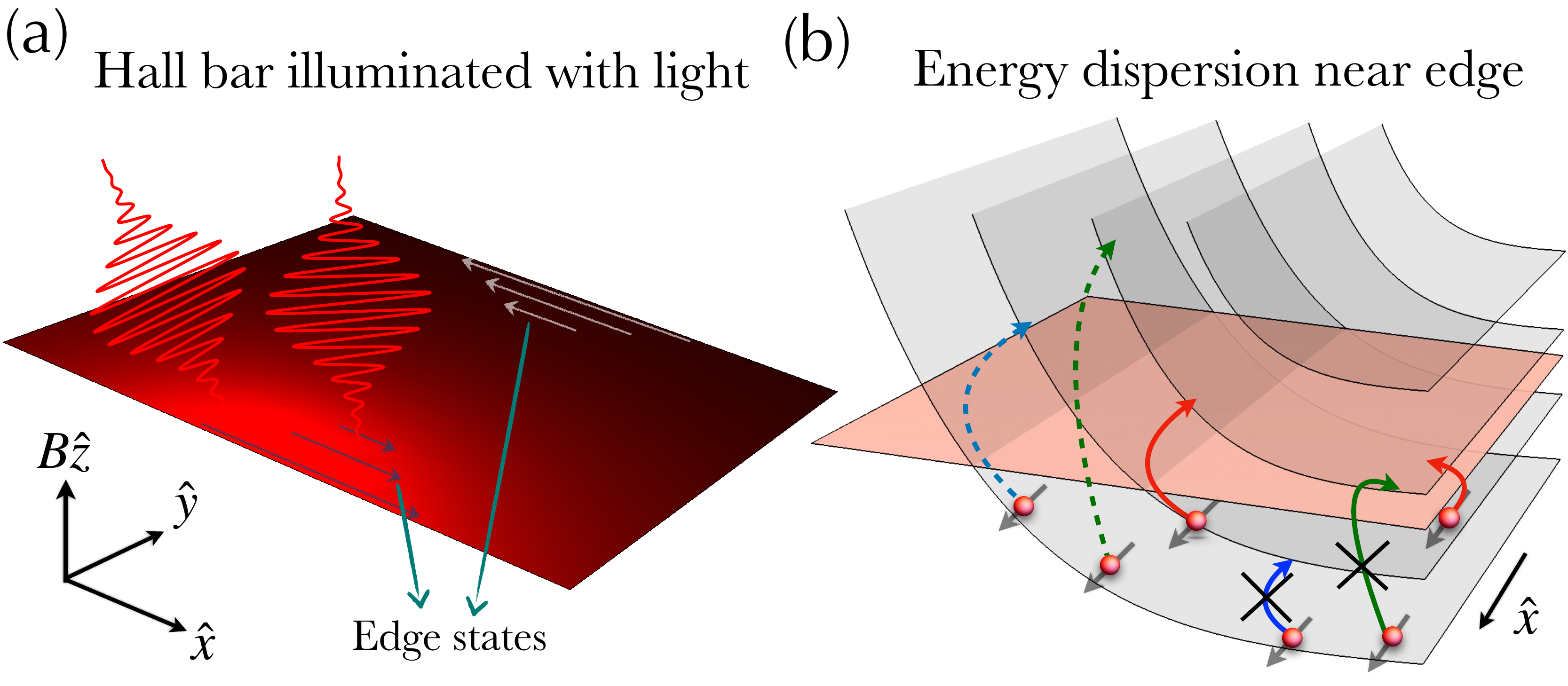}
\caption{(a) Photons interacting with electrons moving near one of the edges of a sample which is subjected to a quantizing magnetic field. (b) Near the edge of the sample, energies and wave-functions of the electrons are modified in such a way that new channels open up for the absorption of photons or nonlinear optical rectification. Dipole forbidden (solid green) and Pauli blocked (solid blue) channels deep in the bulk become active for electrons moving near the edge (marked as dashed green and solid red, respectively).
}\label{fig1a}
\end{figure}

\begin{figure}[!]
\includegraphics[width = \linewidth]{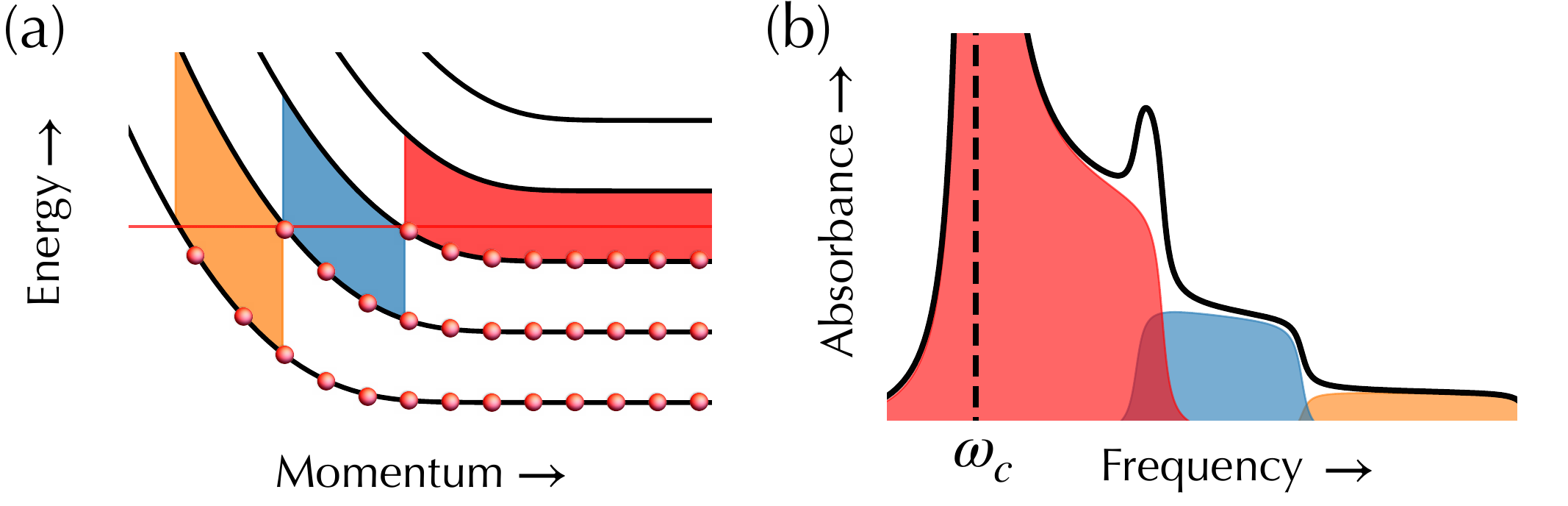}
\caption{(a) The plots of eigenenergies as a function of momentum $k$. Electron states in shaded regions contribute to photon absorption for a given Fermi level shown with red horizontal line.  (b) The absorbance spectrum. Absorption due to the edge states appears at frequencies higher than the bulk cyclotron resonance frequency $\omega_c$. The shaded colored regions under the black curve show the contributions of individual transitions between the neighboring LLs, with matching colors between (a) and (b). 
}\label{fig1b}
\end{figure}

In this paper we show that the optical spectroscopy and even coherent optical control of the QH edge states are still possible and can be in fact very effective without destroying the QH effect conditions in the bulk. The key physical reason for this is that the optical transitions between electron states near the sample boundary (within a few magnetic lengths from the edge) have significantly different transition energies and polarization selection rules as compared to the bulk of the sample. This permits highly selective excitation of a given 1D edge channel with single quasiparticle sensitivity without disturbing the rest of the sample. Furthermore, inversion symmetry breaking near the sample boundary (see highly asymmetric wave functions in the supplemental video) enables strong second-order optical nonlinearity in electric dipole approximation, resulting in efficient optical rectification of incident radiation and direct optical driving of a quasi-DC current in edge states.  These qualitative features are illustrated in Figs.~\ref{fig1a} and \ref{fig1b}, with more quantitative discussion in the sections below. 

High spatiotemporal and energy selectivity of the optical excitations of chiral edge states not only makes it a sensitive spectroscopy tool complementary to electron transport measurements but also enables coherent control of individual edge channels in QH interferometers endowing them with new optoelectronic functionality. Further enhancement in selectivity could be possible with near field tip-enhanced optical microscopy as opposed to far-field illumination sketched in Fig.~\ref{fig1a}. Therefore we hope that our paper will stimulate further collaboration between optical and QH effect communities. 

In this paper we focus at the integer QH effect in semiconductor quantum well samples for parabolic electron dispersion. The graphene edge states offer a greater variety of the optical transitions due to two kinds of edge terminations and will be considered elsewhere, as well as edge states with fractional statistics of carriers. The structure of the paper is as follows. In section \ref{CES} we provide both asymptotic and exact numerical solutions for electron eigenstates and  eigenenergies in the presence of an edge treated as a hard-wall boundary, which of course provides maximum nonadiabaticity. We also calculate dipole matrix elements of the optical transitions between both bulk and edge states. In section \ref{Absorp_prob} we compute a single-photon absorption probability for a quantized optical field and obtain spectra of 2D absorbance which show a series of sharp characteristic peaks at high frequencies that are entirely due to nonadiabatic edge states.  In section \ref{DC_gen}, we demonstrate different optical mechanisms that give rise to direct current (DC) generation by nonlinear rectification of the incident radiation. We calculate the second-order nonlinear DC current which exist already in electric dipole approximation due to inversion symmetry breaking near the edge. We also evaluate the DC current due to optical rectification beyond electric dipole approximation, which exists due to the optical field gradient on the sample.   Appendix summarizes the results for electron eigenstates in the adiabatic approximation for comparison, derives the electron density flux for an arbitrary electric potential, and gives the list of second-order nonlinear density matrix elements contributing to the rectification current. 

%


%

\section{Chiral edge states close to the boundary}\label{CES}
Consider a QH system with electrons constrained to the $x,y$ plane in a quantizing magnetic field $B$ along $z$-axis described by the vector potential ${\vec A} = -y B \hat x$. Near the boundary in $y$ direction located at $y = 0$, the electron states experience the edge potential  $\Phi(y)$, so that  the Schr\"{o}dinger equation for a wave-function of the form $\psi_k(x,y) \sim e^{ikx}\chi(y)$ becomes
\begin{align}\label{parabolic_main}
 \frac{\partial^2\chi}{\partial y^2} + \frac{2m_e}{\hbar^2}\left[E + e\Phi(y) - \frac{1}{2}m_e\omega_c^2\left(y-y_k\right)^2\right]\chi(y) = 0~.
 \end{align}
 where  $m$ is the electron mass, $E$ is the energy eigenvalue, $\omega_c = eB/(m_ec)$ is the cyclotron frequency, $y_k = k\ell_c^2$ is the position of the center of the cyclotron oscillator, and $\ell_c^2 = c\hbar/(eB)$ the magnetic length.

In the adiabatic approximation which is usually invoked when discussing edge states (see Appendix A), $\Phi(y)$ is considered such a slow function of $y$ that its change over the magnetic length scale is neglected.  In this case, while the energies of the LLs increase towards the edge, the energy separation between them does not change and the dipole matrix elements of all transitions between them do not change either. 
Furthermore, the drift velocities of electrons in the channels do not depend on the LL index.

The adiabatic approximation becomes increasingly inadequate when the distance to the edge becomes of the order of a few magnetic lengths. In order to proceed, we need to specify the shape of the edge potential. It obviously depends on the details of the interface and is affected by the presence of the surface states, any space charge accumulation which tends to further sharpen the potential profile \cite{PhysRevB.46.4026}, etc. In order to obtain quantitative results we take the simplest nonadiabatic potential: the hard wall condition  \cite{Mei_1983, Patlatiuk2018}, or a step in  $\Phi(y)$ at $y = 0$, which for a high enough potential barrier means that for a given LL index $n$ and x-component of momentum $k$, $\chi_{nk}(0) = 0$ and $\chi_{nk}(y\to \infty) = 0$. Although obviously an idealization of any realistic boundary, the hard wall condition captures the main physical effects of the nonadiabatic interface, namely asymmetry of the wave functions near the boundary (see the supplemental video), nonequidistant eigenenergies $E_{nk}$ of the edge states as shown in Fig.~\ref{eigenvalues}, modified polarization selection rules which increasingly favor $y$-polarization as the states are pressed to the edge, and inversion symmetry breaking which leads to nonzero permanent dipole moments and modified LL number $n$ selection rules: the transitions with $n$ changing by 2 become allowed in the electric dipole approximation.  Finally, the electron drift velocity in the state $|nk\rangle$ i.e., the diagonal matrix element of the velocity operator, becomes $n$ and $k$ dependent and has a standard form (see Appendix B for derivation):
\begin{align}\label{gr_vel}
\langle \hat v_x \rangle_{nk} = \frac{1}{\hbar}\frac{\partial E_{nk}}{\partial k}~.
\end{align}
Note that this expression  does not depend on the specific form of the eigenstates $\psi_{nk}(x,y)$; moreover, Eq.\eqref{gr_vel} does not depend on any assumptions about the nonuniformity scale of the potential $\Phi(y)$ in comparison with the magnetic length $\ell_c$.

We will consider the observational consequences of these effects one by one below, keeping the derivation details and lengthy formulas in the Appendix.  Before using a numerical solution of Eq.~(\ref{parabolic_main}) with a hard wall potential to calculate the optical response, we point out two approximate analytical solutions of Eq.~(\ref{parabolic_main}) for a hard-wall boundary condition $\Phi(y > 0) = 0$ and $\chi_{nk}(0) = 0$.  


\subsection{Asymptotic solution for $n \gg 1$}

In this case, analytical solutions can be obtained within the quasi-classical approximation \cite{Landau1981}. 
For $y_k > \sqrt{2E/(m \omega_c^2)}$, the quasi-classical solution of Eq.\eqref{parabolic_main} corresponds to two turning points in the region $y>0$ (one at $y > y_k > 0$ and another at $y_k > y > 0$), for which $E = m\omega_c^2\left(y - y_k\right)^2/2$. In this case the solution obeys Bohr-Sommerfeld quantization rule, which for Eq.~\eqref{parabolic_main} corresponds to standard eigenenergies $E_n = \hbar \omega_c (n+1/2) $. 
For $y_k < \sqrt{2E/(m \omega_c^2)}$ there is one turning point at $y > y_k > 0$ and another one due to reflection from the wall at $y = 0$. Denoting the turning point at $y \neq 0$ as $y^*$, we write the quasi-classical solution of Eq.~\eqref{parabolic_main} as 
%
\begin{align}
\chi_{nk}(y) = \frac{1}{N_{nk}\sqrt{q_{nk}}}\cos\left(\int_{y}^{y^*}dy~q_{nk}(y) - \frac{\pi}{4}\right),
\end{align}
where $\hbar q_{nk}(y) = \sqrt{2m\left[E_{nk} - m\omega_c^2(y - y_k)^2/2\right]} = 0$ and $q_{nk}(y^*) = 0$. Here $N_{nk}$ is the normalization factor.
Taking into account the boundary condition on the ideally reflecting wall, i.e., $\chi_{nk}(0) = 0$, we obtain the transcendental equation for the eigenenergies as 
\begin{align}
E_{nk} = \frac{2\pi (n+3/4)\hbar\omega_c}{2\beta_{nk} - \sin(2\beta_{nk})}~, 
\end{align}
with $\cos(\beta_{nk}) = - \hbar k/\sqrt{2mE_{nk}}$.
For $y_k \propto k > 0$ and $1 - \hbar k/\sqrt{2mE_{nk}} \ll 1$, we have $\beta_{nk} \to \pi$, i.e. $E_{nk} = (n+3/4)\hbar\omega_c$. As we see, the energy is increased by $\hbar\omega_c/4$ as compared to the LL in the bulk. This is the result of changing boundary conditions from a smooth effective potential to a hard wall. For $y_k = 0$, $\beta_{nk} = \pi/2$, which gives $E_{nk}=(2n+3/2)\hbar\omega_c$. Here the energy is raised by more than a factor of 2 as compared to the LLs in the bulk.
For $y_k < 0$, one can find the asymptotic solution for $E_{nk} \gg n\hbar\omega_c$, which corresponds to the large wave-number  limit, $|k|\ell_c \gg\sqrt{n}$. In this case we need to have $\beta_{nk} - \sin(2\beta_{nk})/2 \ll 1$, i.e., $\beta_{nk} \to 0$. Denoting $\varepsilon_{k} = \hbar^2k^2/(2m)$, we obtain 
\begin{align}\label{E_quasi}
E_{nk} \approx\varepsilon_{k} + \left[\frac{3\pi}{2}\hbar\omega_c\left(n+\frac{3}{4}\right)\right]^{2/3}\varepsilon_{k}^{1/3}~,
\end{align}
implying that the energies increase and the distance between LLs grows non-equidistantly. This already suggests that for a vertical wall or for any non-adiabatic potential one can realize resonant optical transitions between edge states without causing resonant absorption between bulk LLs.


%
\subsection {Solution for any $n$ but large wavenumbers $k$}

Let us denote $V(y) = m\omega_c^2(y-y_k)^2/2$. Under the condition
$V^{\prime}(y)|_{y=y^*} - V^{\prime}(y)|_{y=0} \ll V^{\prime}(y)|_{y=0}$, where $E = V(y^*)$, 
and for $\xi = \left(2|k|/\ell_c^2\right)^{1/3}\left(y - (E - \varepsilon_{k})/(\hbar\omega_c|k|)\right)$, Eq. \eqref{parabolic_main} 
can be transformed to  
$$ 
\frac{\partial^2\chi}{\partial \xi^2} -\xi\chi = 0.
$$
We can choose our solution as the Airy function ${\rm Ai}(\xi)$, which goes to zero for $\xi\to \infty$ and satisfies the boundary condition $\chi(y=0) =0$, with the corresponding eigenenergies  
\begin{align}\label{E_airy}
~~~~E_{nk} = \varepsilon_{k} + \left(\hbar\omega_c\right)^{2/3}\varepsilon_{k}^{1/3}|\xi_n|~,
\end{align}
where $\xi_n$ are zeros of the Airy function, i.e., ${\rm Ai}(\xi_n) = 0, n = 0,1,2...$. Here $\xi_0$ is the zero with value closest to $\xi = 0$, and their values $|\xi_n|$ increase with increasing $n$. One can verify that Eq.~\eqref{E_airy} is valid as long as $|k|\ell_c \gg \sqrt{n}$. The eigenfunctions $\chi_{nk}(y)$ corresponding to eigenenergies $E_{nk}$ are defined in the interval $0 \leq y < \infty$ as
\begin{align}
~~~~\chi_{nk}(y) = \frac{1}{{\tilde N}_{nk}}{\rm Ai}\left(y\left(\frac{2|k|}{\ell_c^2}\right)^{1/3} + \xi_n\right)~,
\end{align}
with ${\tilde N}_{nk}$ being the normalization factor which is given as, 
\begin{align}
{\tilde N}_{nk}^2 = \left(\frac{\ell_c^2}{2|k|}\right)^{1/3}\int_{\xi_n}^{\infty}{\rm Ai}^2(\xi)d\xi~.
\end{align}
The dispersion of eigenenergies \eqref{E_airy} has the similar structure to those in the previous subsection, even though the former does not rely on $n$ being large.


\begin{figure}[!]
\includegraphics[width = \linewidth]{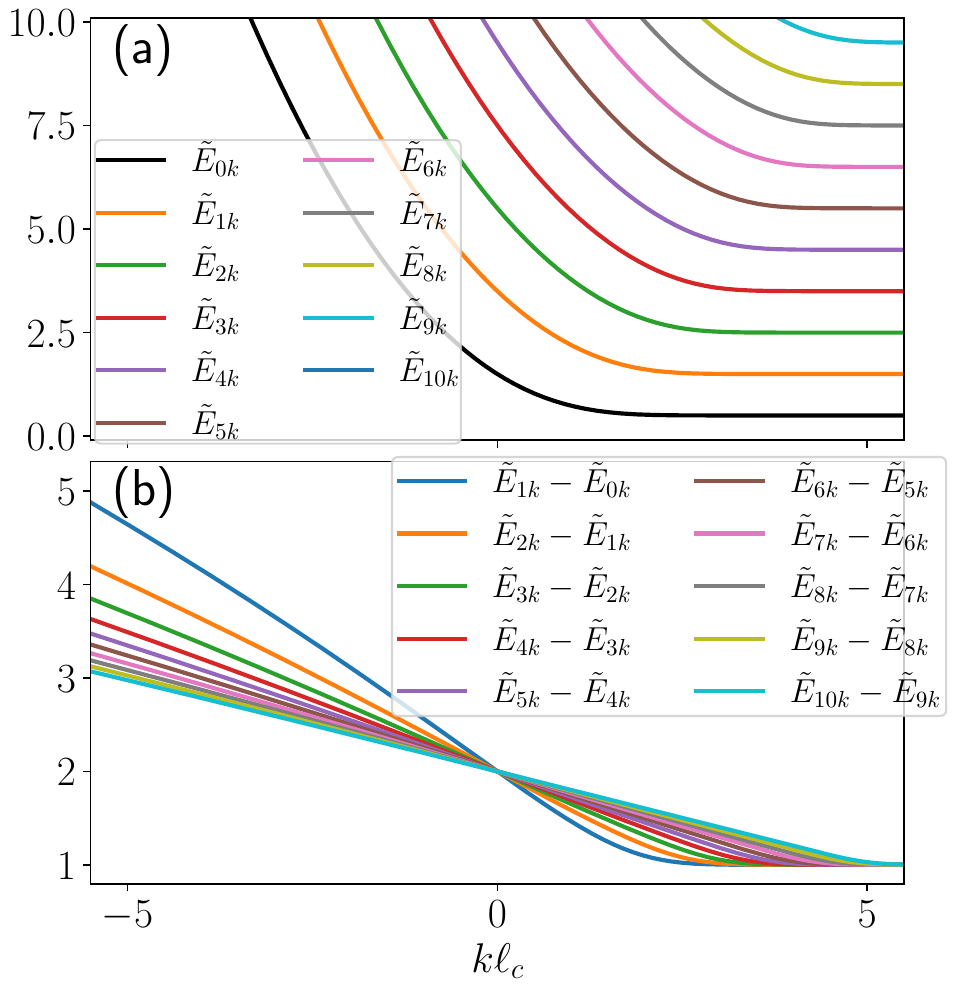}
\caption{(a) Numerically obtained energy eigenvalues (normalized with $\hbar\omega_c$) from Eq.~\eqref{parabolic_main} in the presence of hard wall at $y = 0$ for the first eleven eigenvalues corresponding to $n = 0,1...10$ respectively, as a function of the center of cyclotron rotation $y_k/\ell_c = k\ell_c$. (b) Difference between consecutive energy eigenvalues as a function of $k\ell_c$. Here $\tilde E_{nk}\equiv E_{nk}/\hbar\omega_c$.}
\label{eigenvalues}
\end{figure}


\subsection{Exact numerical solution for eigenstates and dipole matrix elements}

Finally, we compute the eigenvalues and the eigenfunctions for Eq.~\eqref{parabolic_main} numerically using Numerov's algorithm.
Fig.~\ref{eigenvalues} shows the eigenenergies for the first 11 LLs as a function of $y_k/\ell_c = k\ell_c$, and the energy differences between neighboring eigenvalues. It  demonstrates not only the bending of flat LLs of the bulk states near the edge of the sample, but also the fact that the optical transitions become increasingly nonequidistant and move to higher energies as compared to the transitions between bulk LLs. It is also worth pointing out strong inhomogeneous broadening of the inter-LL optical transitions.

\begin{figure}[!]
\includegraphics[width = \linewidth]{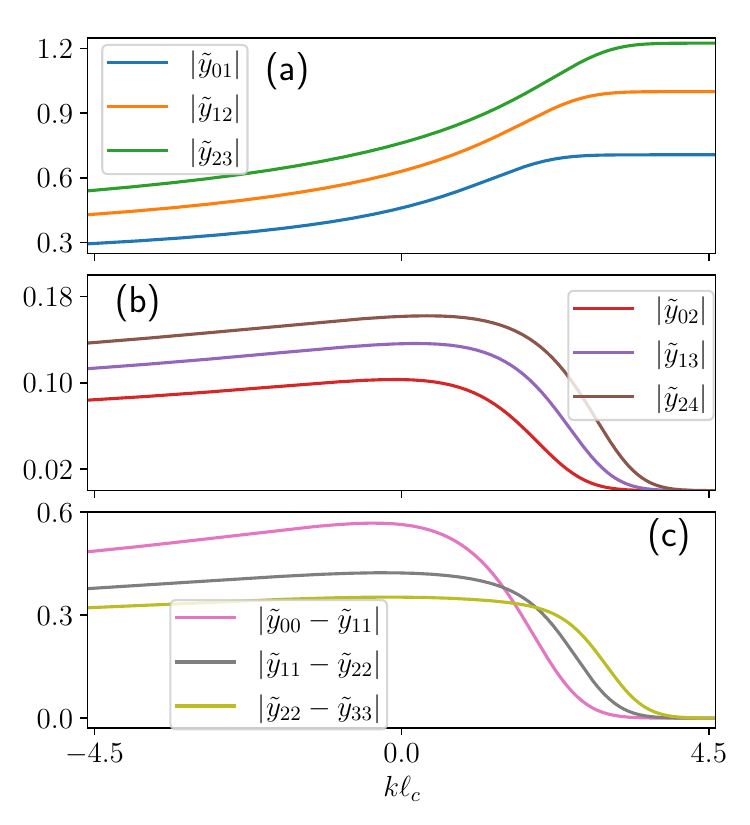}
\caption{Magnitude of the normalized dipole matrix elements as a function of the center of the cyclotron rotation $y_k/\ell_c = k\ell_c$ for (a). $n\to n+1$, (b) $n\to n+2$ and (c) the difference of intra-LL dipole elements. Here $\tilde y_{ij}\equiv y_{ik;jk}/\ell_c$.}
\label{dipole}
\end{figure}

The dipole matrix elements and selection rules are modified drastically in the presence of the boundary, mostly due to inversion symmetry breaking.  In the electric dipole approximation, the general expression for carrier velocity along $\hat x$ direction which includes both diagonal ($n=m$) and off-diagonal ($n\neq m$) elements is 
\begin{align}\label{sel_rule2}
\langle nk|\hat v_x |mk^{\prime}\rangle = \delta_{kk^{\prime}}\omega_c\left(y_k\delta_{nm} - y_{nk;mk} \right)~,
\end{align}
with $y_{nk;mk} = \langle nk|y|mk\rangle$.  
In the bulk, vertical transitions between $m$th and $n$th LLs obeys the following selection rule,
\begin{align}\label{sel_rule1}
\frac{y_{nk;mk}}{\ell_c} = \delta_{n,m+1}\sqrt{\frac{n}{2}}~.
\end{align}
This simple relation, however, does not hold for edge states.
In Fig.~\ref{dipole}, we plot the dipole elements as a function of the center of the cyclotron rotation $y_k/\ell_c = k\ell_c$ for transitions involving the first several LLs. The panel (a) shows that the dipole elements deviate from Eq.~(\ref{sel_rule1}) as $y_k$ approaches the edge. At the same time, previously forbidden transitions $n\to n+2$ are activated, as shown in panel (b).   Furthermore, the permanent, intra-LL dipole matrix elements differ from their bulk values as well (panel (c)), such that their differences between consecutive LLs change from zero to finite values. Both effects are a consequence of inversion symmetry breaking and lead to large second-order nonlinearity in electric-dipole approximation, as we will see below.  

%
\section{Quantum theory of the photon absorption}\label{Absorp_prob}


Since one of the most interesting possibilities offered by the optical field is to excite a single electron into a given edge state, we need a fully quantized theory of fermions in a QH sample interacting with a quantized electromagnetic field. We will derive both the probability of a single photon absorption and the absorbance of a classical  field. 

It is convenient to describe the quantum state of electrons in terms of occupation numbers of $|nk\rangle = \chi_{nk}(y) e^{i k x}$ states, for example $| \cdots 1_{nk} \cdots 0_{n'k'} \cdots \rangle $, where $1_{nk}$ and $0_{n'k'}$  are occupied and unoccupied states.  Fermionic annihilation and creation operators are acting on these states in the usual way: $\hat{a}_{nk} | \cdots 1_{nk} \cdots \rangle = | \cdots 0_{nk} \cdots \rangle$ and $\hat{a}_{nk}^{\dagger} | \cdots 0_{nk} \cdots \rangle = | \cdots 1_{nk} \cdots \rangle$.  The electron Hamiltonian is then $\hat{H}_e = \sum_{n,k}E_{nk}\hat{a}_{nk}^{\dagger}\hat{a}_{nk}$.

The Hamiltonian and the operator of the vector potential of the EM field incident on the sample are 
\begin{equation}
\hat{H}_{\rm ph} = \sum_{\nu,{\bf q}}\hbar\omega_{\nu {\bf q}}\left(\hat{b}_{\nu {\bf q}}^{\dagger}\hat{b}_{\nu {\bf q}} + \frac{1}{2}\right),
\end{equation}
\begin{align}\label{A_field}
\hat {\bf A} = \sum_{\nu, \mathbf{q} }\sqrt{\frac{2\pi c^2\hbar}{V\omega_{\bf q}}}\left({\bf e}_{\nu{\bf q}}\hat{b}_{\nu {\bf q}}e^{i{\bf q}{\bf r}} + {\bf e}^{*}_{\nu{\bf q}}\hat{b}^{\dagger}_{\nu {\bf q}}e^{-i{\bf q}{\bf r}}\right)~,
\end{align}
where $\hat{b}_{\nu \mathbf{q} }$ and $ \hat{b}^{\dagger}_{\nu\mathbf{q} }$ are the photon annihilation  and creation operators acting on the photon Fock states $|n_{\nu \mathbf{q} }\rangle $, $\omega_q = c|{\bf q}|$, $V$ is the quantization volume with periodic boundary conditions,  and ${\bf e}_{\nu{\bf q}}$ is the polarization vector such that ${\bf q}\cdot{\bf e}_{\nu{\bf q}} =0.$
The total Hamiltonian is 
\begin{align}
\hat{H} = \hat{H}_e + \hat{H}_{\rm ph} + \hat{H}_{int}~,
\end{align}
where the interaction Hamiltonian $\hat{H}_{int} = -\frac{1}{c} \hat {\bf j}  \cdot \hat {\bf A} $ and the current operator is $\hat {\bf j}  = - e \hat {\bf v} $. 
From the general relation ${\bf v}_{nk;n'k'} = - i \omega_{nk;n'k'} {\bf r}_{nk;n'k'}$,  where $\hbar\omega_{nk;n'k'} = E_{nk} - E_{n'k'}$,  we can express the current matrix elements through the dipole matrix elements discussed in the previous section: 
\begin{eqnarray} 
{\bf j}_{nk;n'k'} &=& {\bf j}_{nk;n'k}\delta_{kk'}, \\
 (j_x)_{nk;n'k} &=& e \omega_c y_{nk;n'k}, \\ (j_y)_{nk;n'k} &=& i e \omega_{nk;n'k} y_{nk;n'k}, 
\end{eqnarray}
which results in the modified polarization selection rules:
\begin{equation} 
(j_x)_{nk;n'k} + i \frac{\omega_c}{\omega_{nk;n'k}} (j_y)_{nk;n'k} = 0. 
\label{selec}
\end{equation}
As follows from Eq.~(\ref{selec}) and Fig.~\ref{eigenvalues}(b), for electron states near the edge  the transition frequency $\omega_{nk;n'k}$ becomes several times larger than $\omega_c$ and therefore the y-component of the current matrix element becomes significantly larger than the x-component. 

Now consider normal incidence of the radiation on the Hall sample, when ${\bf q}= q {\bf z}_0$, and take the field quantization volume as a ray bundle of volume $V = l_xl_yl_z$. For our purpose it is sufficient to consider only one spatial mode at frequency $\omega_q = cq$ and polarization ${\bf e}_q$. It is straightforward to generalize it to a multimode wave-packet.
As a result, in the rotating-wave approximation (RWA) the interaction Hamiltonian becomes
\begin{align}
\hat{H}_{int} = -\sum_{k,n > n^{\prime}}\sqrt{\frac{2\pi\hbar}{V\omega_{q}}}
\left({\bf e}_{q}\cdot {\bf j}_{nk;n'k}{\hat b}_{q}{\hat a}^{\dagger}_{kn}{\hat a}_{kn^{\prime}} + H.c.\right) ~.\nonumber 
\end{align}
As an initial state of electrons we take
$$\Psi_e (0) = \prod_{\lbrace{nk}\rbrace_{occ}} {\hat a}^{\dagger}_{kn}|0_e\rangle $$
where $|0_e\rangle$ is the vacuum state of the electron system, $\lbrace{nk}\rbrace_{occ}$  denotes occupied states. The values of $n$ change from $n=0$ to $n = n_F$ where $n_F$ is the index of the highest occupied LL in the bulk, whereas the values of $k$ change up to $k_{nF}$, for which $E_{nk}=E_F$, where $E_F$ is the Fermi energy. An initial single-photon state of the field is
$$\Psi_{\rm ph} (0)={\hat b}^{\dagger}_{q}|0_{p}\rangle, $$
where $|0_{p}\rangle$ is the vacuum state of the field. We seek the solution of the Schr\"{o}dinger equation as 
\begin{eqnarray}\label{Wave_func}\nonumber
\Psi(t) &=& C_q(t)\Psi_{\rm ph} (0)\Psi_e (0)\\
&+& \sum_{\lbrace{n^{\prime}k}\rbrace_{\rm occ},n} C_{nn^{\prime}k}(t)|0_{p}\rangle \hat{a}_{nk}^{\dagger}\hat{a}_{n^{\prime}k}\Psi_e (0)~.
\end{eqnarray}
In Eq.~\eqref{Wave_func} the summation is performed only over the indices $n, n^{\prime} $ and $k$ which at the initial moment of time correspond to occupied states $|n^{\prime}k\rangle$ and empty states $|nk\rangle$. At $t=0$ we have $C_q(0) = 1$ and $C_{nn^{\prime}k}(0)=0.$
The Schr\"{o}dinger equation leads to linear equations for the complex amplitudes $C_q(t)$ and $C_{nn^{\prime}k}(t)$, that are similar to those for a quantum field interacting with an inhomogeneously broadened ensemble of two-level quantum emitters \cite{PhysRevA.107.013721}. 
%
%
Solving them, one can obtain a complete quantum dynamics of light-matter interaction. In particular, in the perturbative linear regime we obtain  the absorption probability per unit time and per given transition $n'\to n$ as  
\begin{align}\label{absorp_dim}
{\mathcal A}_{n'\to n} = \sum_k\frac{4\pi|{\bf j}_{nk;n'k}\cdot{\bf e}^{*}_{q}|^2}{\hbar\omega(l_xl_yl_z)}\frac{\Gamma\left(\rho^{(0)}_{n^{\prime}k;n^{\prime}k}-\rho^{(0)}_{nk;nk}\right)}{\Gamma^2 + (\omega_{nk;n'k} -\omega)^2}~,
\end{align}
where $\Gamma$ is the homogeneous broadening of inter-LL transitions determined by disorder-induced scattering and $\rho^{(0)}_{jk;jk} = \left(1+e^{(E_{jk}-E_F)/(k_BT)}\right)^{-1}$ is the Fermi distribution function at temperature $T$. For small $\Gamma$ as compared to the inhomogeneous broadening of transitions between edge states, the Lorentzian in Eq.~\eqref{absorp_dim} becomes the delta function $\pi \delta(\omega_{nk;n'k} -\omega)$ and one recovers the Fermi's golden rule expression. In the case of a finite $\Gamma$ stochastic noise terms appear in the equations for complex amplitudes according to the stochastic Schroedinger equation approach \cite{PhysRevA.107.013721}, but this will not affect the linear absorption probability.  

The summation over $k$ can be replaced by integration, $\sum_k \to g_s \frac{L_x}{2\pi} \int dk$,   
where we also added the spin degeneracy factor $g_s$. Here $L_x$ is the sample length (quantization length of electron states) in x-direction along the edge.  The total absorption probability is obtained by adding the contributions from all LL transitions,  $ {\mathcal A}_{tot} = \frac{l_z}{c} \sum_{n,n^{\prime}}{\mathcal A}_{n^{\prime}\to n}$. Here we also converted the probability per unit time into the total dimensionless probability of absorbing a photon by multiplying the former by the photon pulse duration $\Delta t \approx l_z/c$. Defined this way, the photon absorption probability will also describe the dimensionless absorbance of the classical monochromatic wave by a 2D system. Here we obviously assumed a rectangular temporal profile of the pulse for simplicity. Propagation of single-photon pulses of an arbitrary shape involves a bit more algebra and can be found, e.g., in Ref. \cite{PhysRevLett.131.233802}.

\begin{figure*}[!]
\includegraphics[width = 0.8\linewidth]{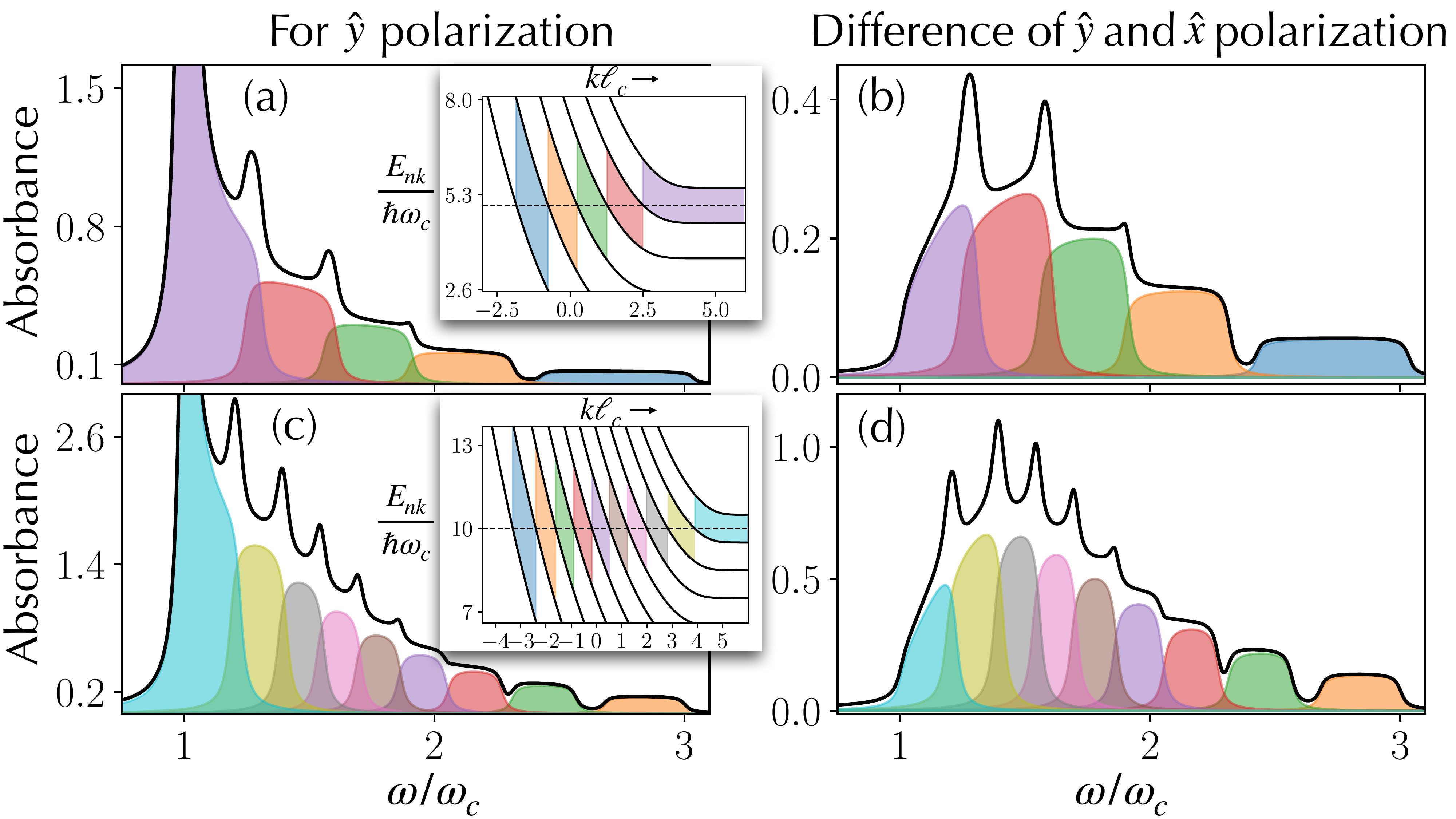}
\caption{Normalized absorbance spectra given by Eq.~\eqref{absor} (solid black line) of $\hat y$-polarized photons for (a) Fermi energy above the fifth LL and (c) Fermi energy above the tenth LL. Panels (b) and (d) show the difference between the absorbance of $\hat y$ and $\hat x$ polarization for the same Fermi levels ($x$-polarized absorbance subtracted from $y$ polarized one). The shaded colored regions under the black curves show the contributions of individual transitions between the neighboring LLs.  The insets in (a) and (c) show the regions of phase space which contribute to the absorption of photons for each pair of neighboring LLs. The colors of shades in the inset match those in the main figure for each panel. Other parameters are  $\Gamma = 0.02\,\omega_c$ and temperature $T = 0$.}
\label{absorp_2deg}
\end{figure*}

The resulting expression depends on the geometric parameter $\frac{L_x}{l_x l_y}$, which can be written as a dimensionless factor $$f = \frac{L_x}{l_x} \frac{ \ell_c}{l_y}.$$ This factor measures the overlap of an incident radiation beam with electron states in the (x,y) plane and is expected to be much smaller than 1, especially due to the ratio $\ell_c/l_y$. Indeed, for a GaAs/AlGaAs quantum well with the conduction band effective mass of $0.067 m_0$ \cite{PhysRevB.43.11787} and in a magnetic field  $B = 1$ T, we have $\hbar\omega_c \approx 1.7$ meV and $\ell_c \approx 26$ nm. If we want to focus incident THz radiation tightly on one edge of the sample in order to bring the overlap closer to 1, one should illuminate a metal tip or fabricate a metallic nanoantenna along the sample edge. 

In order to show universal, geometry-independent absorbance spectra, we normalize the absorbance by this geometric overlap factor, namely  
\begin{equation}
{\mathscr A}(\omega) = \frac{{\mathcal A}_{tot}}{f} = \frac{1}{f}  \frac{l_z}{c} \sum_{n,n^{\prime}}{\mathcal A}_{n^{\prime}\to n}.
    \label{absor} 
\end{equation}

The resulting spectra ${\mathscr A}(\omega)$ of absorbance (or single-photon absorption probability) are shown in Fig.~\ref{absorp_2deg}(a) and (c) for the case of $y$-polarized radiation. The shaded regions under the black curve show the contributions of individual transitions between neighboring LLs; the range of these transitions in momentum space is shown in the insets, with colors matching the shaded regions in the main figure. The salient feature of the total absorbance spectra is the existence of the sharp peaks at frequencies where the contributions from two different transitions overlap. The overlaps exist in the region of momenta $k > 0$ and frequencies $\omega_c < \omega < 2 \omega_c$, as one can also see from the spread of transition energies in Fig.~\ref{eigenvalues}(b). For higher frequencies and for the states closer to the edge ($k < 0$) the overlaps disappear and the peaks in Fig.~\ref{absorp_2deg} turn into dips, as is also clear from Fig.~\ref{eigenvalues}(b) where the order of transition energies between neighboring LLs gets inverted upon crossing $k = 0$. These peaks and dips provide a clear spectroscopic signature of individual edge states. Moreover, they also provide an opportunity of exciting nonequilibrium carriers into a given edge state and then observe a corresponding change in the edge-state interferometry. Due to a large spectral detuning from the bulk transitions at $\omega = \omega_c$ this is possible without destroying the QH effect conditions in the bulk of the sample. 

One detrimental factor that could possibly affect the edge state spectroscopy, especially with broadband pulses, is the presence of a strong peak at $\omega = \omega_c$ due to the bulk state transition $n_F \to n_F + 1$. However, this peak can be subtracted out in the measurements of difference between absorbance in x- and y-polarizations, as shown in Fig.~\ref{absorp_2deg} (b) and (d). Here we make use of the fact that near the edge the x- and y-components of transition matrix elements become increasingly different, with a much stronger y-component, as follows from  Eq.~\eqref{selec}. Thus, difference measurements will get rid of the bulk peak and may result in easier observable peaks and dips for edge states. 

As one can see from Fig.~\ref{absorp_2deg}, the magnitude of absorbance peaks normalized by the geometric factor $f$ is of the order of 1-2. Therefore, the fraction of light absorbed in individual peaks is essentially the above factor $f$ measuring the overlap of the incident radiation with the sample area. While there is no doubt that $f \ll 1$, one needs to keep it large enough to ensure that the absorption is detectable. 


%
%

%

%

%

%
One can also estimate the magnitude of absorbance analytically by taking the limit $\Gamma\to 0$ which converts 
Eq.~\eqref{absorp_dim} into the one given the Fermi's Golden Rule, with the delta-function in the integrand describing energy conservation. Performing integration in $k$, we obtain   
\begin{align}
\frac{l_z}{c} {\mathcal A}_{n^{\prime}\to n} = f \frac{2 \pi|{\bf j}_{nk;n'k}\cdot{\bf e}^{*}_{q}|^2}{\hbar\omega c\ell_c \partial_k\omega_{nk;n^{\prime}k}}\Bigg{|}_{\omega_{nk;n^{\prime}k} = \omega},
\end{align}
where we also took the population difference equal to 1. 
Using the expression for energy dispersion provided in the Eq.~\eqref{E_quasi} (or Eq.~\eqref{E_airy}), we find $\partial_k\omega_{nk;n^{\prime}k} = 2\omega_{nk;n^{\prime}k}/(3k)$. For a $y$-polarized field we have $|{\bf j}_{nk;n'k}\cdot{\bf e}^{*}_{q}|^2 = e^2\omega_{nk;n^{\prime}k}^2|y_{nk;n^{\prime}k}|^2$. This gives 
\begin{align}
\frac{l_z}{c} {\mathcal A}_{n^{\prime}\to n} = f \frac{3 \pi \alpha}{\ell_c} |y_{nk;n^{\prime}k}|^2 k\big{|}_{\omega_{nk;n^{\prime}k} = \omega}~,
\end{align}
where $\alpha$ is the fine structure constant, $k$ scales as $\omega_{nk;n^{\prime}k}^{3/2}$, and the momentum dependence of $|y_{nk;n^{\prime}k}|^2$ can be extracted from Fig.~\ref{dipole}. For example, if we take the wavenumber of the initial state at the Fermi level, $k \approx k_F \approx \frac{1}{\ell_c} \sqrt{2 n_F + 1}$, the absorbance due to the transition between these LLs is   
\begin{align}
\frac{l_z}{c} {\mathcal A}_{n^{\prime}\to n} \approx f 3 \pi \alpha \sqrt{2 n_F + 1} \frac{|y_{nk_F;n^{\prime}k_F}|^2}{\ell_c^2} .
\end{align}

%
\section{Generation of DC current by optical rectification}\label{DC_gen}
The DC current carried by chiral edge states in QH effect experiments is proportional to the diagonal elements of the density matrix, 
\begin{align}
\label{J_diag}
\mathcal{J}_0 = -\frac{e}{L_x} \sum_{n,k}\left(v_x\right)_{nk;nk}\rho^{(0)}_{nk;nk}~. 
\end{align}
The net current from both edges is zero, but it becomes nonzero if a DC voltage $\Delta V$ is applied. Indeed, using Eq.~\eqref{J_diag} together with Eq.~\eqref{gr_vel}, one can find out that in equilibrium, the difference between the currents along opposite edges $\Delta \mathcal{J}_0 = \frac{e^2 n_F}{2 \pi \hbar} \Delta V$, where $n_F$ is the number of filled LLs in the bulk. 

However, inversion symmetry breaking for edge states gives rise to a possibility of generating the DC or quasi-DC current  by an {\it optical} field through the second-order process of the optical rectification, which becomes allowed in the electric dipole approximation and moreover quite efficient. In order to support a net current only one edge needs to be illuminated; otherwise the contributions of two opposite edges will still cancel each other. 

This optically driven DC current is due to the nondiagonal elements of the density matrix,
\begin{align}
\mathcal{J}_{\rm dc} = \frac{e^2B}{L_x m_ec}\sum_{n\neq m} \sum_k y_{nk;mk}\rho_{mk;nk}~.
\label{rect1} 
\end{align}
The density matrix elements are found by solving the master equation,
\begin{align}\label{eqn_of_motion}
\frac{\partial \rho_{nm}}{\partial t} + \frac{i}{\hbar}\sum_{\nu}\left(H_{n\nu}\rho_{\nu m} - \rho_{n\nu}H_{\nu m}\right) = 0~.
\end{align}
to which we will add phenomenological relaxation terms. Here $H_{\alpha\beta} = \delta_{\alpha\beta}E_{\alpha} + V_{\alpha\beta}(t)$ and $\alpha, \beta$ represent the quantum state comprising the Landau level index and the momentum. 
The light-matter interaction Hamiltonian in electric-dipole approximation is $\hat V(t) = ey({\mathcal E}e^{-i\omega t}+{\mathcal E}^*e^{i\omega t})$, where we assume the field to be classical in this section. 
Since all optical transitions are vertical in the electric-dipole approximation, we suppressed the momentum index for simplicity, such that $y_{ik;jk}\to y_{ij}$ and $\rho_{ik;jk}\to \rho_{ij}$. Of course the momentum index has to be restored when integrating over $k$ in Eq.~\eqref{rect1}.  

We proceed by expanding the elements of the density matrix in perturbative series with respect to the interaction Hamiltonian, 
%
%
\begin{align}
\frac{\partial \rho^{(j+1)}_{nm}}{\partial t} =  - i\omega_{nm}{\rho}^{(j+1)}_{nm} -\frac{i}{\hbar}\sum_{\nu}\left(V_{n\nu}\rho^{(j)}_{\nu m} - \rho^{(j)}_{n\nu}V_{\nu m}\right),
\end{align}
with $n$ and $m$ denoting the quantum states corresponding to Landau level indices $n$ and $m$ respectively. We are only interested in the terms up to the second order in the optical field. The linear in ${\mathcal E}$ term yields 
\begin{align}\label{coherence_alpha_beta}
{\rho}^{(1)}_{\beta\alpha} =  -\frac{e}{\hbar}\frac{y_{\beta\alpha}\left(\rho^{(0)}_{\alpha\alpha} - \rho^{(0)}_{\beta\beta}\right)}{\omega_{\beta\alpha} - \omega - i\gamma}~\mathcal{E}e^{-i\omega t}~,
\end{align}
where ${\rho}^{(1)}_{\alpha\beta} = \left({\rho}^{(1)}_{\beta\alpha}\right)^{*}$.

From Eq.~\eqref{rect1} and taking into account the structure of the perturbative solution, it is clear that there are two types of contributions to the DC current at zero frequency, i.e., $\propto |{\mathcal E}|^2$. One type of terms in Eq.~\eqref{rect1} scale as  $y_{n;(n+2)}\rho^{(2)}_{(n+2);n}$. They are enabled by nonzero dipole matrix elements $y_{n;(n+2)}$ of the transitions which change the LL index by 2, i.e., $n \to n+2 $. The second group of terms scale as $y_{n;(n+1)}\rho^{(2)}_{(n+1);n}$ where $ \rho^{(2)}_{(n+1);n} \propto (y_{(n+1);(n+1)} - y_{n;n})$.  As one can see from Fig.~\ref{dipole}, both $y_{n;(n+2)}$ and $(y_{(n+1);(n+1)} - y_{n;n})$ are only nonzero for electron states close to the edge, when the center of the cyclotron rotation $y_k$ is within several magnetic lengths $\ell_c$ from the edge. 

\begin{figure}[!]
\includegraphics[width = \linewidth]{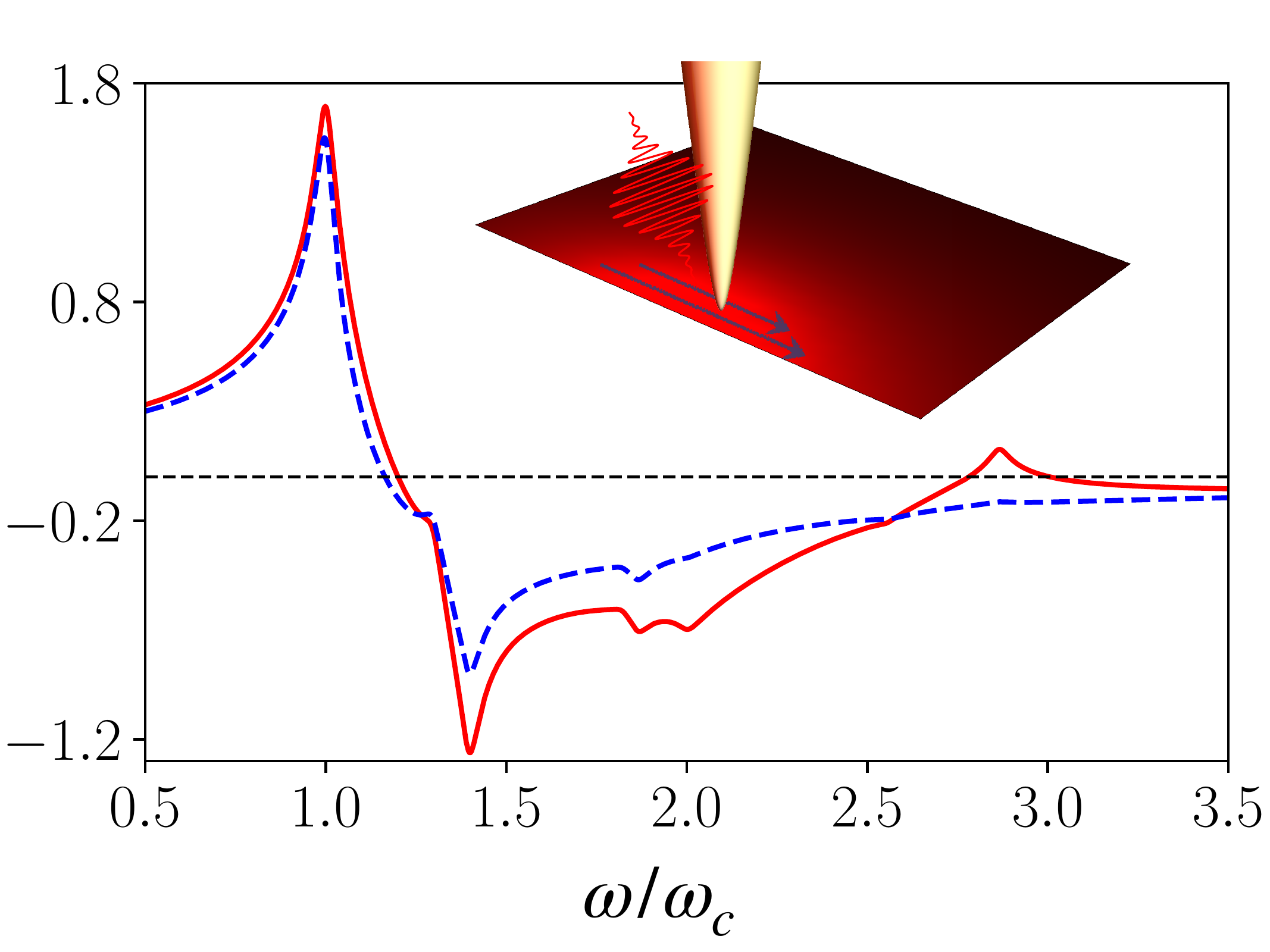}
\caption{Nonlinear DC current resulting from second-order optical rectification process, as a function of the normalized optical field frequency. The calculations included the first five Landau levels for chemical potential $E_F = 3\hbar\omega_c$. Red curve: the electric dipole contribution to the nonlinear current from  Eq.~\eqref{J12}, $\frac{{\mathcal J}^{(I)}_{\rm dc} + {\mathcal J}^{(II)}_{\rm dc}}{\zeta \mathcal{J}_0}$, normalized by the ``diagonal'' DC current $\mathcal{J}_0$  from Eq.~\eqref{J_diag} and factor $\zeta$ from Eq.~(\ref{zeta}). Blue dashed curve: beyond electric-dipole contribution to the nonlinear current, $\frac{{\mathcal J}^{(III)}_{\rm dc} + {\mathcal J}^{(IV)}_{\rm dc}}{\zeta_{\rm grad} \mathcal{J}_0}$, normalized by $\mathcal{J}_0$ and factor $\zeta_{\rm grad}$ from Eq.~(\ref{grad}).  }
\label{non_linear}
\end{figure}

The second-order nonlinear density matrix elements contributing to the DC current satisfy the equations of motion,
  \begin{eqnarray}\nonumber
		&&\frac{\partial \rho^{(2)}_{n;(n-1)}}{\partial t} + i\omega_{n;(n-1)}\rho^{(2)}_{n;(n-1)} = -\frac{ie }{\hbar} {\mathscr F}(t) \times\\\nonumber
		&&\left[ y_{n;(n-2)}\rho^{(1)}_{(n-2);(n-1)} - y_{(n-2);(n-1)}\rho^{(1)}_{n;(n-2)}\right.\\\nonumber
		&&+y_{n;(n+1)}\rho^{(1)}_{(n+1);(n-1)} - y_{(n+1);(n-1)}\rho^{(1)}_{n;(n+1)}\\
		&&\left.+(y_{n;n} - y_{(n-1);(n-1)})\rho^{(1)}_{n;(n-1)}\right],
\end{eqnarray}
  \begin{eqnarray}\nonumber
		&&\frac{\partial \rho^{(2)}_{(n+1);(n-1)}}{\partial t} + i\omega_{(n+1);(n-1)}\rho^{(2)}_{(n+1);(n-1)} = \\\nonumber
		&&-\frac{ie}{\hbar}{\mathscr F}(t) \left[ y_{(n+1);n}\rho^{(1)}_{n;(n-1)} - y_{n;(n-1)}\rho^{(1)}_{(n+1);n}\right.\\
		&&\left.+(y_{(n+1);(n+1)} - y_{(n-1);(n-1)})\rho^{(1)}_{(n+1);(n-1)}\right],
\end{eqnarray}
where we have defined ${\mathscr F}(t) = {\mathcal E}e^{-i\omega t}+{\mathcal E}^*e^{i\omega t}$.
Time-independent solution to these equations under illumination with a monochromatic field gives rise to a large number of terms contributing to the DC current, which grows rapidly with increasing doping. For convenience, in Appendix C below we list the density matrix elements that contribute to the rectification current for the Fermi level $E_F = 3\hbar\omega_c$ between $n = 2$ and $n = 3$ bulk LLs, chosen to make plots in Fig.~\ref{non_linear}. 
In Fig.~\ref{non_linear} the red curve shows the rectification current normalized by the conventional ``diagonal'' DC current (\ref{J_diag}) and by the factor 
\begin{equation}
\label{zeta}
\zeta = \frac{e^2|{\mathcal E}|^2\ell_c^2}{\hbar^2 \omega_c^2} 
\end{equation}
proportional to the optical field intensity, as a function of the optical field frequency. The current is a sum of the above two contributions, ${\mathcal J}^{(I)}_{\rm dc}$ and ${\mathcal J}^{(II)}_{\rm dc}$ 
given by
\begin{align}\label{J12}
{\mathcal J}^{(I/II)}_{\rm dc} = \frac{e^2B}{m_ec}\int \frac{dk}{2\pi}~J^{(I/II)}~,
\end{align}
where 
\begin{eqnarray}
J^{(I)}\label{J_1} &=& y_{01}{\rho}^{(2)}_{10} + y_{12}{\rho}^{(2)}_{21} + y_{23}{\rho}^{(2)}_{32} + c.c.~,\\
J^{(II)}\label{J_2} &=& y_{02}{\rho}^{(2)}_{20} + y_{13}{\rho}^{(2)}_{31} + y_{24}{\rho}^{(2)}_{42} + c.c~.
\end{eqnarray}
The expressions for the second-order density matrix elements are given in Appendix C. 

The main peak at the cyclotron frequency in Fig.~\ref{non_linear} is due to the $n = 2 \to n = 3$ transition near the edge. With increasing optical frequency, the term $(E_{3k}-E_{2k})/\hbar - \omega$ flips sign and consequently we observe a negative peak in the plot. Additional kinks are due to activation/deactivation of various transitions. They are going to be smeared by finite temperature effects and/or increased scattering. 

Since the normalized nonlinear current in Fig.~\ref{non_linear} is of the order of 1, the dimensional current magnitude is fraction $\zeta$ of the conventional edge  current (\ref{J_diag}). The factor $\zeta$ is essentially the Rabi frequency squared of incident light normalized to the cyclotron frequency squared. When $\zeta$ becomes of the order of 1, population transfer effects between the LLs in the bulk become important even despite the detuning from the bulk inter-LL resonance. This could create bulk conductance across the sample and destroy the QH effect. Therefore, to maintain the QH effect conditions in the bulk, it is desirable to keep $\zeta \ll 1$. The optimal situation to observe or utilize the nonlinear rectification current would be to illuminate one edge of the sample under zero DC bias conditions, when the  conventional current is zero. 


\subsection{Beyond electric-dipole approximation}

Yet another second-order nonlinear optical contribution to the rectified DC current originates from the spatial nonuniformity of the electric field of the incident radiation, i.e., beyond the electric-dipole approximation. In the lowest order, the light-mater interaction Hamiltonian becomes 
$$ \hat V(t) = e\left[ \left( y{\mathcal E} + \frac{y^2}{2} \partial_y {\mathcal E} \right) e^{-i\omega t}+\left( y{\mathcal E} + \frac{y^2}{2} \partial_y {\mathcal E} \right)^*e^{i\omega t} \right]. $$
This gives rise to the following additional matrix elements contributing to the rectification current, 
$$  
V_{(n+2);n} = \frac{e}{2} (y^2)_{(n+2);n} \left( \partial_y {\mathcal E}  e^{-i\omega t}+ \partial_y {\mathcal E}^* e^{i\omega t} \right),
$$
$$  
V_{(n+1);(n-1)} = \frac{e}{2} (y^2)_{(n+1);(n-1)} \left( \partial_y {\mathcal E}  e^{-i\omega t}+ \partial_y {\mathcal E}^* e^{i\omega t} \right).
$$
Solving the density matrix equations in the second order of the field as in the previous subsection,  we obtain additional contributions to the density matrix elements. For example the factor $y_{12}y_{20}{\mathcal E}^*{\mathcal E}$ in the first term in ${\rho}^{(2)}_{10}$ in Appendix C is modified as $y_{12}y_{20}{\mathcal E}^*{\mathcal E} + y_{12} y^2_{20}{\mathcal E}^*\partial_y{\mathcal E}/2  + y^2_{12}y_{20}  {\mathcal E}  \partial_y{\mathcal E} ^*/2$, and similarly for other terms.

%

We denote the resulting contributions as $J^{III}$ and $J^{IV}$ in the spirit of Eq.~\eqref{J_1} and Eq.~\eqref{J_2}, respectively, and calculate the resulting rectification current ${\mathcal J}^{(III)}_{\rm dc} + {\mathcal J}^{(IV)}_{\rm dc}$ similarly to  Eq.~\eqref{J12}. It is shown as a blue curve in Fig.~\ref{non_linear}. The current is normalized by the conventional ``diagonal'' DC current (\ref{J_diag}) and by the factor 
\begin{equation} 
\label{grad} 
\zeta_{\rm grad} = \frac{e^2 \ell_c^3 \partial_y |{\mathcal E}|^2}{\hbar^2 \omega_c^2} 
\end{equation}
proportional to the gradient of the optical field intensity. With this normalization, the optical rectification current is close in magnitude to the one obtained in electric dipole approximation and shown as a red curve. Note, however, that the normalization factor in the denominator contains the optical field intensity gradient $ \ell_c \partial_y |{\mathcal E}|^2$ which is much smaller than $|{\mathcal E}|^2$, unless the optical field is focused to the size of the order of  $\ell_c$, i.e., tens of nm, by using a nanotip or nanoantenna at the sample edge as sketched in the inset to Fig.~\ref{non_linear}. Without nanofocusing, the electric-dipole current is greater in magnitude by the factor $\sim \frac{|{\mathcal E}|^2}{\ell_c \partial_y |{\mathcal E}|^2}$.   


\section{Conclusions}

In conclusion, 
we demonstrated the feasibility of the optical spectroscopy and selective coherent optical control of chiral edge state populations and current under the conditions of the integer QH effect.  The physical mechanism enabling the selective control is inversion symmetry breaking and violation of adiabaticity for electron states near the edge. As a result, optical transitions between the edge states have significantly different transition energies and polarization selection rules as compared to the bulk of the sample. This enables  selective excitation of a given edge channel with single quasiparticle sensitivity without disturbing the rest of the sample. The 2D absorbance shows characteristic peaks at different edge state resonances at frequencies significantly higher than the bulk cyclotron transition. A large fraction of incident light in the illuminated area of the sample can be absorbed. The overall absorption probability of single photons depends on their overlap with the edge state area and will benefit from nanofocusing with a tip or nanoantenna.  Furthermore, inversion symmetry breaking near the sample boundary enables strong second-order optical nonlinearity already in electric dipole approximation, resulting in efficient optical rectification of incident radiation and direct optical driving of a quasi-DC current in edge states. The predicted optical effects can be used to to study or control edge currents by optical means and to control the interference pattern in QH interference experiments. 

While the calculations in this paper were performed for nonrelativistic electron dispersion in semiconductor quantum wells, one should expect qualitatively similar results in graphene samples. Another natural extension of this work is to investigate the optical control of edge current under fractional QH effect conditions. We hope that this study will attract attention of both QH and optical communities and lead to interesting collaborative experiments.

%
\section{Acknowledgements} 
This work has been supported in part by the Air Force Office for Scientific Research Grant No. FA9550-21-1-0272 and
National Science Foundation Award No. 1936276. 

%
\appendix
%


\section{Electron states in adiabatic approximation}

For the reader's convenience, here we summarize the results for electron states in the adiabatic approximation, when the electric potential $\Phi(y)$ varies much slower than the magnetic length.  
For a magnetic field, ${\vec B} = B \hat z$ we choose the vector potential of the form ${\vec A} = -y B \hat x$, which yields the single-electron Hamiltonian 
\begin{align}
H =  \frac{1}{2m}\left(p_x - \frac{eBy}{c}\right)^2 + \frac{p_y^2}{2m} - e\Phi~,
\end{align}
where $m$ is the electron mass. Since there is no explicit $x$ dependence, we can seek the wavefunction as  
\begin{align}\label{wave_func}
\psi(x,y) = e^{ikx}\chi(y)~, 
\end{align}
for which the Schroedinger equation yields 
 \begin{align}
 \frac{\partial^2\chi}{\partial y^2} + \frac{2m}{\hbar^2}\left[E + e\Phi(y) - \frac{1}{2}m\omega_c^2\left(y-y_k\right)^2\right]\chi(y) = 0~.
 \end{align}
where we have defined the cyclotron frequency, $\omega_c = eB/(mc)$, the position of the center of the cyclotron rotation $y_k = k\ell_c^2$, and the magnetic length $\ell_c = \sqrt{c\hbar/(eB)}$.

In the adiabatic approximation, the potential $\Phi(y)$ is varying slowly near the edge, $\Phi(y) \approx \Phi(y_k) + \left(y-y_k\right)\left(\partial_y\Phi(y)|_{y = y_k}\right)$. The particle energies are  
\begin{eqnarray}\nonumber
E_{nk} &=& \left(n+\frac{1}{2}\right)\hbar\omega_c - e\Phi\left(y_k + \frac{c\partial_y\Phi(y)|_{y = y_k}}{\omega_c B}\right)\\
&+& \frac{m}{2}\left(\frac{c\partial_y\Phi(y)|_{y = y_k}}{B}\right)^2. 
\label{adiab}
\end{eqnarray}
The term $\frac{c \partial_y\Phi(y)|_{y = y_k}}{\omega_c B}$ in the right-hand side of Eq.~\eqref{adiab} is the shift of the center of the cyclotron rotation with respect to the initial point $y_k$ under the action of the electric field  $E_y \approx - \partial_y\Phi(y)|_{y = y_k}$. The last term on the right-hand side of Eq.~\eqref{adiab} is the kinetic energy of the drift motion in crossed $E$ and $B$ fields. 

The eigenfunctions become 
\begin{eqnarray}\nonumber
\chi_{nk}(y) &=& \frac{e^{-\left(y-y_k-c\partial_y\Phi(y)|_{y = y_k}/(\omega_c B)\right)^2/(\sqrt{2}\ell_c)^2}}{\sqrt{2^nn! \sqrt{\pi}\ell_c}}\\
&\times&{\mathcal H}_n\left(\frac{y}{\ell_c}-\frac{y_k}{\ell_c}-\frac{c}{\ell_c}\frac{\partial_y\Phi(y)|_{y = y_k}}{\omega_c B}\right), 
\label{chi_adiab}
\end{eqnarray}
where $n$ is the LL index and ${\mathcal H}_n$ is a Hermite polynomial.  In this case, while the energies of the LLs increase towards the edge, the energy separation between them does not change as compared to the bulk states, and the dipole matrix elements of all transitions between them do not change either. 
Furthermore, the drift velocity of electrons along the edge does not depend on the LL index:
\begin{align}
\langle \hat v_x \rangle_{k} = -\frac{c}{B}\partial_y\Phi(y)|_{y = y_k}~.
\end{align}
%


\section{Electron density flux for an arbitrary potential}
It is interesting that the electron density flux and drift velocity can be found for  an arbitrary potential $\Phi(y)$, provided  Eq.~\eqref{parabolic_main} allows a finite solution corresponding to a certain discrete spectrum of $E_{nk}$ and a set of eigenfunctions of the form 
$\psi_{nk}(x,y) = e^{ikx}\chi_{nk}(y)$, where 
$\chi_{nk}(y)|_{y\to \pm\infty}\to 0$, $\int dxdy~\psi_{nk}^*(x,y)\psi_{mk^{\prime}}(x,y) = \delta_{nm}\delta_{kk^{\prime}}$.
To prove this, we introduce the operator of two-dimensional electron density for the $n$th LL, ${\hat N}_{n}(x,y) = \hat{\Psi}^{\dagger}(x,y)\hat{\Psi}(x,y)$ where the field operators are given in the second-quantized form:
$\hat{\Psi}(x,y) = \sum_k \psi_{nk}(x,y){\hat a}_{nk}, \hat{\Psi}^{\dagger} = \sum_k \psi^*_{nk}(x,y){\hat a}^{\dagger}_{nk}$ with the annihilation and creation operators as $\hat a_{nk}$ and $\hat{a}^{\dagger}_{nk}$ respectively.
The operator of two-dimensional spatial electron density is then 
\begin{align}
{\hat N}_{n}(x,y) = \sum_{k,k^{\prime}} \chi^*_{nk^{\prime}}(y)\chi_{nk}(y)e^{i(k-k^{\prime})x}\hat\rho_{nk;nk^{\prime}}~,
\end{align}
where $\hat\rho_{nk;nk^{\prime}} = {\hat a}^{\dagger}_{nk}{\hat a}_{nk^{\prime}}$. The same result can be obtained by using the transition operator $\hat\rho_{nk;nk^{\prime}}= |nk\rangle\langle nk^{\prime}|$. The operator $\hat\rho_{nk;nk^{\prime}}$ obeys the Heisenberg equation,
\begin{align}\label{den_nn}
\frac{\partial\hat\rho_{nk;nk^{\prime}}}{\partial t} = -\frac{i}{\hbar}\left(E_{nk} - E_{nk^{\prime}}\right)\hat\rho_{nk;nk^{\prime}}~.
\end{align}
If the fermions are interacting with classical electromagnetic fields, the Heisenberg operator $\hat\rho_{nk;nk^{\prime}}$, after averaging over the initial state of the system, becomes the matrix element of a standard density matrix $\rho_{nk;nk^{\prime}}$.
The general expression for the probability flux of the electron density can be expressed through the Wigner function, defined as 
\begin{align} {\hat {\mathcal W}}_{n}(x,y,K) = \nonumber \\ 
 \sum_{\kappa}\psi^*_{n(K-\kappa/2)}(x,y)\psi_{n(K+\kappa/2)}(x,y)\hat\rho_{n(K+\kappa/2);n(K-\kappa/2)}, \nonumber 
\end{align}
where $k-k^{\prime} = \kappa$, $(k+k^{\prime})/2 = K$. The Wigner function ${\hat {\mathcal W}}_{n}(x,y,K)$ determines both the operator of spatial density, 
\begin{align} \label{spat} 
\sum_K {\hat {\mathcal W}}_{n}(x,y,K) = \hat N_{n}(x,y),
\end{align}
and the distribution over momenta, 
\begin{align} \label{moment} 
\int dx dy~{\hat {\mathcal W}}_{n}(x,y,K)|_{K = k} = \hat\rho_{nk;nk}.
\end{align}
It follows from Eq.~\eqref{den_nn} that 
\begin{align} \label{den2}
\frac{\partial\hat\rho_{n(K+\kappa/2);n(K-\kappa/2)}}{\partial t} = \nonumber \\
-\frac{i}{\hbar}\left(E_{n(K+\kappa/2)} - E_{n(K-\kappa/2)}\right)\hat\rho_{n(K+\kappa/2);n(K-\kappa/2)}~.
\end{align}

Suppose that we have a well-localized state in $k$ so that $\delta k \ll k$ (a narrow spectrum in $k$ means narrow localization along $y$ near the edge). In this case we have $E_{n(K+\kappa/2)}-E_{n(K-\kappa/2)}\approx \kappa\partial_kE_{nk}|_{k = K}$. Then from Eq.~\eqref{den2}  and taking into account Eqs.~\eqref{spat} and \eqref{moment} one can obtain  
\begin{align}\label{W_xy}
\frac{\partial {\hat {\mathcal W}}_{n}(x,y,K)}{\partial t} +  \frac{1}{\hbar}\frac{\partial E_{nk}}{\partial k}\bigg{|}_{k=K}\frac{\partial {\hat  {\mathcal W}}_{n}(x,y,K)}{\partial x} = 0~.
\end{align}
For a quadratic spectrum of the type $E_{nk}=E_{n0}+\alpha_nk^2$ Eq.~\eqref{W_xy} is exact and does not require a narrow spectrum. 

Using Eqs.~\eqref{W_xy} and \eqref{spat}, one can obtain 
\begin{align}\label{N_xy}
\frac{\partial {\hat N}_{n}(x,y)}{\partial t} +  \frac{1}{\hbar}\frac{\partial E_{nk}}{\partial k}\frac{\partial {\hat N}_{n}(x,y)}{\partial x} = 0
\end{align}
where $k$ is the central value of the given narrow spectrum. 
 From Eq.\eqref{N_xy} one obtains the expression for the observed particle velocity in the state $|nk\rangle$ i.e., the diagonal matrix element of the velocity operator, which has a standard form:
\begin{align}\label{gr_vel2}
\langle \hat v_x \rangle_{nk} = \frac{1}{\hbar}\frac{\partial E_{nk}}{\partial k}~.
\end{align}
Note that this expression coincides with the one obtained from the solution \eqref{chi_adiab} for a uniform electric field but it does not depend on the specific form of the eigenstates $\psi_{nk}(x,y)$; furthermore, Eq.~\eqref{gr_vel2} does not depend on any assumptions about the nonuniformity scale of the potential $\Phi(y)$ in comparison with the magnetic length $\ell_c$.
%

%
%


%
\section{Second-order nonlinear density matrix elements}\label{DC_gen1}

%
For reader's convenience, we provide below the the list of second-order nonlinear density matrix elements that contribute to the optical rectification current in the electric-dipole approximation for our chosen value of the Fermi level, $E_F = 3\hbar\omega_c$: 
\onecolumngrid 
\begin{eqnarray} 
		\frac{{\rho}^{(2)}_{10}}{e^2|\mathcal{E}|^2/\hbar^2} &=&  
		\frac{y_{12}y_{20}\Delta\rho^{(0)}_{02}}{\omega_{10}\left(\Delta_{20} - i\gamma\right)}
		- \frac{y_{20}y_{12}\Delta\rho^{(0)}_{12}}{\omega_{10}\left(\Delta_{21} + i\gamma\right)} 
- \frac{\left(y_{00}-y_{11}\right)y_{10}\Delta\rho^{(0)}_{01}}{\omega_{10}\left(\Delta_{10} - i\gamma\right)},  \nonumber \\  
\frac{{\rho}^{(2)}_{21}}{e^2|\mathcal{E}|^2/\hbar^2} &=&  \frac{y_{20}y_{01}\Delta\rho^{(0)}_{01}}{\omega_{21}\left(\Delta_{10} + i\gamma\right)}
		-\frac{y_{01}y_{20}\Delta\rho^{(0)}_{02}}{\omega_{21}\left(\Delta_{20} - i\gamma\right)} 
		+\frac{y_{23}y_{31}\Delta\rho^{(0)}_{13}}{\omega_{21}\left(\Delta_{31} - i\gamma\right)}
		-\frac{y_{31}y_{23}\Delta\rho^{(0)}_{23}}{\omega_{21}\left(\Delta_{32} + i\gamma\right)} 
-\frac{\left(y_{11}-y_{22}\right)y_{21}\Delta\rho^{(0)}_{12}}{\omega_{21}\left(\Delta_{21} - i\gamma\right)}, \nonumber\\
\frac{{\rho}^{(2)}_{32}}{e^2|\mathcal{E}|^2/\hbar^2} &=&  \frac{y_{31}y_{12}\Delta\rho^{(0)}_{12}}{\omega_{32}\left(\Delta_{21} + i\gamma\right)}
		-\frac{y_{12}y_{31}\Delta\rho^{(0)}_{13}}{\omega_{32}\left(\Delta_{31} - i\gamma\right)}
		+\frac{y_{34}y_{42}\Delta\rho^{(0)}_{24}}{\omega_{32}\left(\Delta_{42} - i\gamma\right)}
		-\frac{y_{42}y_{34}\Delta\rho^{(0)}_{34}}{\omega_{32}\left(\Delta_{43} + i\gamma\right)}
-\frac{\left(y_{22}-y_{33}\right)y_{32}\Delta\rho^{(0)}_{23}}{\omega_{32}\left(\Delta_{32} - i\gamma\right)}, \nonumber \\
\frac{{\rho}^{(2)}_{43}}{e^2|\mathcal{E}|^2/\hbar^2} &=&  \frac{y_{42}y_{23}\Delta\rho^{(0)}_{23}}{\omega_{43}\left(\Delta_{32} + i\gamma\right)}
		-\frac{y_{23}y_{42}\Delta\rho^{(0)}_{24}}{\omega_{43}\left(\Delta_{42} - i\gamma\right)}, \nonumber
\\ 
		\frac{{\rho}^{(2)}_{20}}{e^2|\mathcal{E}|^2/\hbar^2} &=& \frac{y_{21}y_{10}\Delta\rho^{(0)}_{01}}{\omega_{20}\left(\Delta_{10} - i\gamma\right)}
		-\frac{y_{10}y_{21}\Delta\rho^{(0)}_{12}}{\omega_{20}\left(\Delta_{21} - i\gamma\right)}
-
\frac{\left(y_{00}-y_{22}\right)y_{20}\Delta\rho^{(0)}_{02}}{\omega_{20}\left(\Delta_{20} - i\gamma\right)},  \nonumber \\
\nonumber
\frac{{\rho}^{(2)}_{31}}{e^2|\mathcal{E}|^2/\hbar^2} &=& \frac{y_{32}y_{21}\Delta\rho^{(0)}_{12}}{\omega_{31}\left(\Delta_{21} - i\gamma\right)}
		-\frac{y_{21}y_{32}\Delta\rho^{(0)}_{23}}{\omega_{31}\left(\Delta_{32} - i\gamma\right)}
-
\frac{\left(y_{11}-y_{33}\right)y_{31}\Delta\rho^{(0)}_{13}}{\omega_{31}\left(\Delta_{31} - i\gamma\right)},  \\
\nonumber
\frac{{\rho}^{(2)}_{42}}{e^2|\mathcal{E}|^2/\hbar^2} &=& \frac{y_{43}y_{32}\Delta\rho^{(0)}_{23}}{\omega_{42}\left(\Delta_{32} - i\gamma\right)}
-
\frac{\left(y_{22}-y_{44}\right)y_{42}\Delta\rho^{(0)}_{24}}{\omega_{42}\left(\Delta_{42} - i\gamma\right)}. \nonumber 
	\end{eqnarray}
Here $\Delta\rho^{(0)}_{\alpha\beta} = \left(\rho^{(0)}_{\alpha\alpha} - \rho^{(0)}_{\beta\beta}\right)$ is the equilibrium population difference and $\Delta_{\alpha\beta} = \omega_{\alpha\beta} - \omega$. 

\twocolumngrid


\bibliography{Ref2}

\begin{thebibliography}{29}%
\makeatletter
\providecommand \@ifxundefined [1]{%
 \@ifx{#1\undefined}
}%
\providecommand \@ifnum [1]{%
 \ifnum #1\expandafter \@firstoftwo
 \else \expandafter \@secondoftwo
 \fi
}%
\providecommand \@ifx [1]{%
 \ifx #1\expandafter \@firstoftwo
 \else \expandafter \@secondoftwo
 \fi
}%
\providecommand \natexlab [1]{#1}%
\providecommand \enquote  [1]{``#1''}%
\providecommand \bibnamefont  [1]{#1}%
\providecommand \bibfnamefont [1]{#1}%
\providecommand \citenamefont [1]{#1}%
\providecommand \href@noop [0]{\@secondoftwo}%
\providecommand \href [0]{\begingroup \@sanitize@url \@href}%
\providecommand \@href[1]{\@@startlink{#1}\@@href}%
\providecommand \@@href[1]{\endgroup#1\@@endlink}%
\providecommand \@sanitize@url [0]{\catcode `\\12\catcode `\$12\catcode
  `\&12\catcode `\#12\catcode `\^12\catcode `\_12\catcode `\%12\relax}%
\providecommand \@@startlink[1]{}%
\providecommand \@@endlink[0]{}%
\providecommand \url  [0]{\begingroup\@sanitize@url \@url }%
\providecommand \@url [1]{\endgroup\@href {#1}{\urlprefix }}%
\providecommand \urlprefix  [0]{URL }%
\providecommand \Eprint [0]{\href }%
\providecommand \doibase [0]{http://dx.doi.org/}%
\providecommand \selectlanguage [0]{\@gobble}%
\providecommand \bibinfo  [0]{\@secondoftwo}%
\providecommand \bibfield  [0]{\@secondoftwo}%
\providecommand \translation [1]{[#1]}%
\providecommand \BibitemOpen [0]{}%
\providecommand \bibitemStop [0]{}%
\providecommand \bibitemNoStop [0]{.\EOS\space}%
\providecommand \EOS [0]{\spacefactor3000\relax}%
\providecommand \BibitemShut  [1]{\csname bibitem#1\endcsname}%
\let\auto@bib@innerbib\@empty
\bibitem [{\citenamefont {Prange}\ and\ \citenamefont {Girvin}(1987)}]{QHE}%
  \BibitemOpen
  \bibinfo {editor} {\bibfnamefont {R.~E.}\ \bibnamefont {Prange}}\ and\
  \bibinfo {editor} {\bibfnamefont {S.~M.}\ \bibnamefont {Girvin}},\ eds.,\
  \href {\doibase 10.1007/978-1-4684-0499-9} {\emph {\bibinfo {title} {The
  Quantum Hall Effect}}}\ (\bibinfo  {publisher} {Springer US},\ \bibinfo
  {year} {1987})\BibitemShut {NoStop}%
\bibitem [{\citenamefont {Halperin}(1982)}]{Halperin_Edge}%
  \BibitemOpen
  \bibfield  {author} {\bibinfo {author} {\bibfnamefont {B.~I.}\ \bibnamefont
  {Halperin}},\ }\href {\doibase 10.1103/PhysRevB.25.2185} {\bibfield
  {journal} {\bibinfo  {journal} {Phys. Rev. B}\ }\textbf {\bibinfo {volume}
  {25}},\ \bibinfo {pages} {2185} (\bibinfo {year} {1982})}\BibitemShut
  {NoStop}%
\bibitem [{\citenamefont {MacDonald}\ and\ \citenamefont
  {St\ifmmode~\check{r}\else \v{r}\fi{}eda}(1984)}]{MacDonald_edge}%
  \BibitemOpen
  \bibfield  {author} {\bibinfo {author} {\bibfnamefont {A.~H.}\ \bibnamefont
  {MacDonald}}\ and\ \bibinfo {author} {\bibfnamefont {P.}~\bibnamefont
  {St\ifmmode~\check{r}\else \v{r}\fi{}eda}},\ }\href {\doibase
  10.1103/PhysRevB.29.1616} {\bibfield  {journal} {\bibinfo  {journal} {Phys.
  Rev. B}\ }\textbf {\bibinfo {volume} {29}},\ \bibinfo {pages} {1616}
  (\bibinfo {year} {1984})}\BibitemShut {NoStop}%
\bibitem [{\citenamefont {Pascher}\ \emph {et~al.}(2014)\citenamefont
  {Pascher}, \citenamefont {R\"ossler}, \citenamefont {Ihn}, \citenamefont
  {Ensslin}, \citenamefont {Reichl},\ and\ \citenamefont
  {Wegscheider}}]{PhysRevX.4.011014}%
  \BibitemOpen
  \bibfield  {author} {\bibinfo {author} {\bibfnamefont {N.}~\bibnamefont
  {Pascher}}, \bibinfo {author} {\bibfnamefont {C.}~\bibnamefont {R\"ossler}},
  \bibinfo {author} {\bibfnamefont {T.}~\bibnamefont {Ihn}}, \bibinfo {author}
  {\bibfnamefont {K.}~\bibnamefont {Ensslin}}, \bibinfo {author} {\bibfnamefont
  {C.}~\bibnamefont {Reichl}}, \ and\ \bibinfo {author} {\bibfnamefont
  {W.}~\bibnamefont {Wegscheider}},\ }\href {\doibase
  10.1103/PhysRevX.4.011014} {\bibfield  {journal} {\bibinfo  {journal} {Phys.
  Rev. X}\ }\textbf {\bibinfo {volume} {4}},\ \bibinfo {pages} {011014}
  (\bibinfo {year} {2014})}\BibitemShut {NoStop}%
\bibitem [{\citenamefont {Marguerite}\ \emph {et~al.}(2019)\citenamefont
  {Marguerite}, \citenamefont {Birkbeck}, \citenamefont {Aharon-Steinberg},
  \citenamefont {Halbertal}, \citenamefont {Bagani}, \citenamefont {Marcus},
  \citenamefont {Myasoedov}, \citenamefont {Geim}, \citenamefont {Perello},\
  and\ \citenamefont {Zeldov}}]{Marguerite2019}%
  \BibitemOpen
  \bibfield  {author} {\bibinfo {author} {\bibfnamefont {A.}~\bibnamefont
  {Marguerite}}, \bibinfo {author} {\bibfnamefont {J.}~\bibnamefont
  {Birkbeck}}, \bibinfo {author} {\bibfnamefont {A.}~\bibnamefont
  {Aharon-Steinberg}}, \bibinfo {author} {\bibfnamefont {D.}~\bibnamefont
  {Halbertal}}, \bibinfo {author} {\bibfnamefont {K.}~\bibnamefont {Bagani}},
  \bibinfo {author} {\bibfnamefont {I.}~\bibnamefont {Marcus}}, \bibinfo
  {author} {\bibfnamefont {Y.}~\bibnamefont {Myasoedov}}, \bibinfo {author}
  {\bibfnamefont {A.~K.}\ \bibnamefont {Geim}}, \bibinfo {author}
  {\bibfnamefont {D.~J.}\ \bibnamefont {Perello}}, \ and\ \bibinfo {author}
  {\bibfnamefont {E.}~\bibnamefont {Zeldov}},\ }\href {\doibase
  10.1038/s41586-019-1704-3} {\bibfield  {journal} {\bibinfo  {journal}
  {Nature}\ }\textbf {\bibinfo {volume} {575}},\ \bibinfo {pages} {628}
  (\bibinfo {year} {2019})}\BibitemShut {NoStop}%
\bibitem [{\citenamefont {Johnsen}\ \emph {et~al.}(2023)\citenamefont
  {Johnsen}, \citenamefont {Schattauer}, \citenamefont {Samaddar},
  \citenamefont {Weston}, \citenamefont {Hamer}, \citenamefont {Watanabe},
  \citenamefont {Taniguchi}, \citenamefont {Gorbachev}, \citenamefont
  {Libisch},\ and\ \citenamefont {Morgenstern}}]{PhysRevB.107.115426}%
  \BibitemOpen
  \bibfield  {author} {\bibinfo {author} {\bibfnamefont {T.}~\bibnamefont
  {Johnsen}}, \bibinfo {author} {\bibfnamefont {C.}~\bibnamefont {Schattauer}},
  \bibinfo {author} {\bibfnamefont {S.}~\bibnamefont {Samaddar}}, \bibinfo
  {author} {\bibfnamefont {A.}~\bibnamefont {Weston}}, \bibinfo {author}
  {\bibfnamefont {M.~J.}\ \bibnamefont {Hamer}}, \bibinfo {author}
  {\bibfnamefont {K.}~\bibnamefont {Watanabe}}, \bibinfo {author}
  {\bibfnamefont {T.}~\bibnamefont {Taniguchi}}, \bibinfo {author}
  {\bibfnamefont {R.}~\bibnamefont {Gorbachev}}, \bibinfo {author}
  {\bibfnamefont {F.}~\bibnamefont {Libisch}}, \ and\ \bibinfo {author}
  {\bibfnamefont {M.}~\bibnamefont {Morgenstern}},\ }\href {\doibase
  10.1103/PhysRevB.107.115426} {\bibfield  {journal} {\bibinfo  {journal}
  {Phys. Rev. B}\ }\textbf {\bibinfo {volume} {107}},\ \bibinfo {pages}
  {115426} (\bibinfo {year} {2023})}\BibitemShut {NoStop}%
\bibitem [{\citenamefont {Li}\ \emph {et~al.}(2013)\citenamefont {Li},
  \citenamefont {Luican-Mayer}, \citenamefont {Abanin}, \citenamefont
  {Levitov},\ and\ \citenamefont {Andrei}}]{Li2013}%
  \BibitemOpen
  \bibfield  {author} {\bibinfo {author} {\bibfnamefont {G.}~\bibnamefont
  {Li}}, \bibinfo {author} {\bibfnamefont {A.}~\bibnamefont {Luican-Mayer}},
  \bibinfo {author} {\bibfnamefont {D.}~\bibnamefont {Abanin}}, \bibinfo
  {author} {\bibfnamefont {L.}~\bibnamefont {Levitov}}, \ and\ \bibinfo
  {author} {\bibfnamefont {E.~Y.}\ \bibnamefont {Andrei}},\ }\href {\doibase
  10.1038/ncomms2767} {\bibfield  {journal} {\bibinfo  {journal} {Nature
  Communications}\ }\textbf {\bibinfo {volume} {4}},\ \bibinfo {pages} {1744}
  (\bibinfo {year} {2013})}\BibitemShut {NoStop}%
\bibitem [{\citenamefont {Patlatiuk}\ \emph {et~al.}(2018)\citenamefont
  {Patlatiuk}, \citenamefont {Scheller}, \citenamefont {Hill}, \citenamefont
  {Tserkovnyak}, \citenamefont {Barak}, \citenamefont {Yacoby}, \citenamefont
  {Pfeiffer}, \citenamefont {West},\ and\ \citenamefont
  {Zumbühl}}]{Patlatiuk2018}%
  \BibitemOpen
  \bibfield  {author} {\bibinfo {author} {\bibfnamefont {T.}~\bibnamefont
  {Patlatiuk}}, \bibinfo {author} {\bibfnamefont {C.~P.}\ \bibnamefont
  {Scheller}}, \bibinfo {author} {\bibfnamefont {D.}~\bibnamefont {Hill}},
  \bibinfo {author} {\bibfnamefont {Y.}~\bibnamefont {Tserkovnyak}}, \bibinfo
  {author} {\bibfnamefont {G.}~\bibnamefont {Barak}}, \bibinfo {author}
  {\bibfnamefont {A.}~\bibnamefont {Yacoby}}, \bibinfo {author} {\bibfnamefont
  {L.~N.}\ \bibnamefont {Pfeiffer}}, \bibinfo {author} {\bibfnamefont {K.~W.}\
  \bibnamefont {West}}, \ and\ \bibinfo {author} {\bibfnamefont {D.~M.}\
  \bibnamefont {Zumbühl}},\ }\href {\doibase 10.1038/s41467-018-06025-3}
  {\bibfield  {journal} {\bibinfo  {journal} {Nature Communications}\ }\textbf
  {\bibinfo {volume} {9}},\ \bibinfo {pages} {3692} (\bibinfo {year}
  {2018})}\BibitemShut {NoStop}%
\bibitem [{\citenamefont {Kim}\ \emph {et~al.}(2021)\citenamefont {Kim},
  \citenamefont {Schwenk}, \citenamefont {Walkup}, \citenamefont {Zeng},
  \citenamefont {Ghahari}, \citenamefont {Le}, \citenamefont {Slot},
  \citenamefont {Berwanger}, \citenamefont {Blankenship}, \citenamefont
  {Watanabe}, \citenamefont {Taniguchi}, \citenamefont {Giessibl},
  \citenamefont {Zhitenev}, \citenamefont {Dean},\ and\ \citenamefont
  {Stroscio}}]{Kim2021}%
  \BibitemOpen
  \bibfield  {author} {\bibinfo {author} {\bibfnamefont {S.}~\bibnamefont
  {Kim}}, \bibinfo {author} {\bibfnamefont {J.}~\bibnamefont {Schwenk}},
  \bibinfo {author} {\bibfnamefont {D.}~\bibnamefont {Walkup}}, \bibinfo
  {author} {\bibfnamefont {Y.}~\bibnamefont {Zeng}}, \bibinfo {author}
  {\bibfnamefont {F.}~\bibnamefont {Ghahari}}, \bibinfo {author} {\bibfnamefont
  {S.~T.}\ \bibnamefont {Le}}, \bibinfo {author} {\bibfnamefont {M.~R.}\
  \bibnamefont {Slot}}, \bibinfo {author} {\bibfnamefont {J.}~\bibnamefont
  {Berwanger}}, \bibinfo {author} {\bibfnamefont {S.~R.}\ \bibnamefont
  {Blankenship}}, \bibinfo {author} {\bibfnamefont {K.}~\bibnamefont
  {Watanabe}}, \bibinfo {author} {\bibfnamefont {T.}~\bibnamefont {Taniguchi}},
  \bibinfo {author} {\bibfnamefont {F.~J.}\ \bibnamefont {Giessibl}}, \bibinfo
  {author} {\bibfnamefont {N.~B.}\ \bibnamefont {Zhitenev}}, \bibinfo {author}
  {\bibfnamefont {C.~R.}\ \bibnamefont {Dean}}, \ and\ \bibinfo {author}
  {\bibfnamefont {J.~A.}\ \bibnamefont {Stroscio}},\ }\href {\doibase
  10.1038/s41467-021-22886-7} {\bibfield  {journal} {\bibinfo  {journal}
  {Nature Communications}\ }\textbf {\bibinfo {volume} {12}},\ \bibinfo {pages}
  {2852} (\bibinfo {year} {2021})}\BibitemShut {NoStop}%
\bibitem [{\citenamefont {van Wees}\ \emph {et~al.}(1989)\citenamefont {van
  Wees}, \citenamefont {Kouwenhoven}, \citenamefont {Harmans}, \citenamefont
  {Williamson}, \citenamefont {Timmering}, \citenamefont {Broekaart},
  \citenamefont {Foxon},\ and\ \citenamefont {Harris}}]{PhysRevLett.62.2523}%
  \BibitemOpen
  \bibfield  {author} {\bibinfo {author} {\bibfnamefont {B.~J.}\ \bibnamefont
  {van Wees}}, \bibinfo {author} {\bibfnamefont {L.~P.}\ \bibnamefont
  {Kouwenhoven}}, \bibinfo {author} {\bibfnamefont {C.~J. P.~M.}\ \bibnamefont
  {Harmans}}, \bibinfo {author} {\bibfnamefont {J.~G.}\ \bibnamefont
  {Williamson}}, \bibinfo {author} {\bibfnamefont {C.~E.}\ \bibnamefont
  {Timmering}}, \bibinfo {author} {\bibfnamefont {M.~E.~I.}\ \bibnamefont
  {Broekaart}}, \bibinfo {author} {\bibfnamefont {C.~T.}\ \bibnamefont
  {Foxon}}, \ and\ \bibinfo {author} {\bibfnamefont {J.~J.}\ \bibnamefont
  {Harris}},\ }\href {\doibase 10.1103/PhysRevLett.62.2523} {\bibfield
  {journal} {\bibinfo  {journal} {Phys. Rev. Lett.}\ }\textbf {\bibinfo
  {volume} {62}},\ \bibinfo {pages} {2523} (\bibinfo {year}
  {1989})}\BibitemShut {NoStop}%
\bibitem [{\citenamefont {de~C.~Chamon}\ \emph {et~al.}(1997)\citenamefont
  {de~C.~Chamon}, \citenamefont {Freed}, \citenamefont {Kivelson},
  \citenamefont {Sondhi},\ and\ \citenamefont {Wen}}]{Chamon}%
  \BibitemOpen
  \bibfield  {author} {\bibinfo {author} {\bibfnamefont {C.}~\bibnamefont
  {de~C.~Chamon}}, \bibinfo {author} {\bibfnamefont {D.~E.}\ \bibnamefont
  {Freed}}, \bibinfo {author} {\bibfnamefont {S.~A.}\ \bibnamefont {Kivelson}},
  \bibinfo {author} {\bibfnamefont {S.~L.}\ \bibnamefont {Sondhi}}, \ and\
  \bibinfo {author} {\bibfnamefont {X.~G.}\ \bibnamefont {Wen}},\ }\href
  {\doibase 10.1103/PhysRevB.55.2331} {\bibfield  {journal} {\bibinfo
  {journal} {Physical Review B}\ }\textbf {\bibinfo {volume} {55}},\ \bibinfo
  {pages} {2331} (\bibinfo {year} {1997})}\BibitemShut {NoStop}%
\bibitem [{\citenamefont {Ji}\ \emph {et~al.}(2003)\citenamefont {Ji},
  \citenamefont {Chung}, \citenamefont {Sprinzak}, \citenamefont {Heiblum},
  \citenamefont {Mahalu},\ and\ \citenamefont {Shtrikman}}]{Ji2003}%
  \BibitemOpen
  \bibfield  {author} {\bibinfo {author} {\bibfnamefont {Y.}~\bibnamefont
  {Ji}}, \bibinfo {author} {\bibfnamefont {Y.}~\bibnamefont {Chung}}, \bibinfo
  {author} {\bibfnamefont {D.}~\bibnamefont {Sprinzak}}, \bibinfo {author}
  {\bibfnamefont {M.}~\bibnamefont {Heiblum}}, \bibinfo {author} {\bibfnamefont
  {D.}~\bibnamefont {Mahalu}}, \ and\ \bibinfo {author} {\bibfnamefont
  {H.}~\bibnamefont {Shtrikman}},\ }\href {\doibase 10.1038/nature01503}
  {\bibfield  {journal} {\bibinfo  {journal} {Nature}\ }\textbf {\bibinfo
  {volume} {422}},\ \bibinfo {pages} {415} (\bibinfo {year}
  {2003})}\BibitemShut {NoStop}%
\bibitem [{\citenamefont {Chung}\ and\ \citenamefont
  {Stone}(2006)}]{PhysRevB.73.245311}%
  \BibitemOpen
  \bibfield  {author} {\bibinfo {author} {\bibfnamefont {S.~B.}\ \bibnamefont
  {Chung}}\ and\ \bibinfo {author} {\bibfnamefont {M.}~\bibnamefont {Stone}},\
  }\href {\doibase 10.1103/PhysRevB.73.245311} {\bibfield  {journal} {\bibinfo
  {journal} {Phys. Rev. B}\ }\textbf {\bibinfo {volume} {73}},\ \bibinfo
  {pages} {245311} (\bibinfo {year} {2006})}\BibitemShut {NoStop}%
\bibitem [{\citenamefont {Stern}\ and\ \citenamefont
  {Halperin}(2006)}]{PhysRevLett.96.016802}%
  \BibitemOpen
  \bibfield  {author} {\bibinfo {author} {\bibfnamefont {A.}~\bibnamefont
  {Stern}}\ and\ \bibinfo {author} {\bibfnamefont {B.~I.}\ \bibnamefont
  {Halperin}},\ }\href {\doibase 10.1103/PhysRevLett.96.016802} {\bibfield
  {journal} {\bibinfo  {journal} {Phys. Rev. Lett.}\ }\textbf {\bibinfo
  {volume} {96}},\ \bibinfo {pages} {016802} (\bibinfo {year}
  {2006})}\BibitemShut {NoStop}%
\bibitem [{\citenamefont {Feldman}\ and\ \citenamefont
  {Kitaev}(2006)}]{PhysRevLett.97.186803}%
  \BibitemOpen
  \bibfield  {author} {\bibinfo {author} {\bibfnamefont {D.~E.}\ \bibnamefont
  {Feldman}}\ and\ \bibinfo {author} {\bibfnamefont {A.}~\bibnamefont
  {Kitaev}},\ }\href {\doibase 10.1103/PhysRevLett.97.186803} {\bibfield
  {journal} {\bibinfo  {journal} {Phys. Rev. Lett.}\ }\textbf {\bibinfo
  {volume} {97}},\ \bibinfo {pages} {186803} (\bibinfo {year}
  {2006})}\BibitemShut {NoStop}%
\bibitem [{\citenamefont {Zhang}\ \emph {et~al.}(2009)\citenamefont {Zhang},
  \citenamefont {McClure}, \citenamefont {Levenson-Falk}, \citenamefont
  {Marcus}, \citenamefont {Pfeiffer},\ and\ \citenamefont
  {West}}]{PhysRevB.79.241304}%
  \BibitemOpen
  \bibfield  {author} {\bibinfo {author} {\bibfnamefont {Y.}~\bibnamefont
  {Zhang}}, \bibinfo {author} {\bibfnamefont {D.~T.}\ \bibnamefont {McClure}},
  \bibinfo {author} {\bibfnamefont {E.~M.}\ \bibnamefont {Levenson-Falk}},
  \bibinfo {author} {\bibfnamefont {C.~M.}\ \bibnamefont {Marcus}}, \bibinfo
  {author} {\bibfnamefont {L.~N.}\ \bibnamefont {Pfeiffer}}, \ and\ \bibinfo
  {author} {\bibfnamefont {K.~W.}\ \bibnamefont {West}},\ }\href {\doibase
  10.1103/PhysRevB.79.241304} {\bibfield  {journal} {\bibinfo  {journal} {Phys.
  Rev. B}\ }\textbf {\bibinfo {volume} {79}},\ \bibinfo {pages} {241304}
  (\bibinfo {year} {2009})}\BibitemShut {NoStop}%
\bibitem [{\citenamefont {Stern}\ \emph {et~al.}(2010)\citenamefont {Stern},
  \citenamefont {Rosenow}, \citenamefont {Ilan},\ and\ \citenamefont
  {Halperin}}]{PhysRevB.82.085321}%
  \BibitemOpen
  \bibfield  {author} {\bibinfo {author} {\bibfnamefont {A.}~\bibnamefont
  {Stern}}, \bibinfo {author} {\bibfnamefont {B.}~\bibnamefont {Rosenow}},
  \bibinfo {author} {\bibfnamefont {R.}~\bibnamefont {Ilan}}, \ and\ \bibinfo
  {author} {\bibfnamefont {B.~I.}\ \bibnamefont {Halperin}},\ }\href {\doibase
  10.1103/PhysRevB.82.085321} {\bibfield  {journal} {\bibinfo  {journal} {Phys.
  Rev. B}\ }\textbf {\bibinfo {volume} {82}},\ \bibinfo {pages} {085321}
  (\bibinfo {year} {2010})}\BibitemShut {NoStop}%
\bibitem [{\citenamefont {Déprez}\ \emph {et~al.}(2021)\citenamefont
  {Déprez}, \citenamefont {Veyrat}, \citenamefont {Vignaud}, \citenamefont
  {Nayak}, \citenamefont {Watanabe}, \citenamefont {Taniguchi}, \citenamefont
  {Gay}, \citenamefont {Sellier},\ and\ \citenamefont
  {Sacépé}}]{Corentin2021}%
  \BibitemOpen
  \bibfield  {author} {\bibinfo {author} {\bibfnamefont {C.}~\bibnamefont
  {Déprez}}, \bibinfo {author} {\bibfnamefont {L.}~\bibnamefont {Veyrat}},
  \bibinfo {author} {\bibfnamefont {H.}~\bibnamefont {Vignaud}}, \bibinfo
  {author} {\bibfnamefont {G.}~\bibnamefont {Nayak}}, \bibinfo {author}
  {\bibfnamefont {K.}~\bibnamefont {Watanabe}}, \bibinfo {author}
  {\bibfnamefont {T.}~\bibnamefont {Taniguchi}}, \bibinfo {author}
  {\bibfnamefont {F.}~\bibnamefont {Gay}}, \bibinfo {author} {\bibfnamefont
  {H.}~\bibnamefont {Sellier}}, \ and\ \bibinfo {author} {\bibfnamefont
  {B.}~\bibnamefont {Sacépé}},\ }\href {\doibase 10.1038/s41565-021-00847-x}
  {\bibfield  {journal} {\bibinfo  {journal} {Nature Nanotechnology}\ }\textbf
  {\bibinfo {volume} {16}},\ \bibinfo {pages} {555} (\bibinfo {year}
  {2021})}\BibitemShut {NoStop}%
\bibitem [{\citenamefont {Feldman}\ and\ \citenamefont
  {Halperin}(2022)}]{PhysRevB.105.165310}%
  \BibitemOpen
  \bibfield  {author} {\bibinfo {author} {\bibfnamefont {D.~E.}\ \bibnamefont
  {Feldman}}\ and\ \bibinfo {author} {\bibfnamefont {B.~I.}\ \bibnamefont
  {Halperin}},\ }\href {\doibase 10.1103/PhysRevB.105.165310} {\bibfield
  {journal} {\bibinfo  {journal} {Phys. Rev. B}\ }\textbf {\bibinfo {volume}
  {105}},\ \bibinfo {pages} {165310} (\bibinfo {year} {2022})}\BibitemShut
  {NoStop}%
\bibitem [{\citenamefont {Nakamura}\ \emph {et~al.}(2020)\citenamefont
  {Nakamura}, \citenamefont {Liang}, \citenamefont {Gardner},\ and\
  \citenamefont {Manfra}}]{Nakamura2020}%
  \BibitemOpen
  \bibfield  {author} {\bibinfo {author} {\bibfnamefont {J.}~\bibnamefont
  {Nakamura}}, \bibinfo {author} {\bibfnamefont {S.}~\bibnamefont {Liang}},
  \bibinfo {author} {\bibfnamefont {G.~C.}\ \bibnamefont {Gardner}}, \ and\
  \bibinfo {author} {\bibfnamefont {M.~J.}\ \bibnamefont {Manfra}},\ }\href
  {\doibase 10.1038/s41567-020-1019-1} {\bibfield  {journal} {\bibinfo
  {journal} {Nature Physics}\ }\textbf {\bibinfo {volume} {16}},\ \bibinfo
  {pages} {931–936} (\bibinfo {year} {2020})}\BibitemShut {NoStop}%
\bibitem [{\citenamefont {Ronen}\ \emph {et~al.}(2021)\citenamefont {Ronen},
  \citenamefont {Werkmeister}, \citenamefont {Haie~Najafabadi}, \citenamefont
  {Pierce}, \citenamefont {Anderson}, \citenamefont {Shin}, \citenamefont
  {Lee}, \citenamefont {Lee}, \citenamefont {Johnson}, \citenamefont
  {Watanabe}, \citenamefont {Taniguchi}, \citenamefont {Yacoby},\ and\
  \citenamefont {Kim}}]{Ron2021}%
  \BibitemOpen
  \bibfield  {author} {\bibinfo {author} {\bibfnamefont {Y.}~\bibnamefont
  {Ronen}}, \bibinfo {author} {\bibfnamefont {T.}~\bibnamefont {Werkmeister}},
  \bibinfo {author} {\bibfnamefont {D.}~\bibnamefont {Haie~Najafabadi}},
  \bibinfo {author} {\bibfnamefont {A.~T.}\ \bibnamefont {Pierce}}, \bibinfo
  {author} {\bibfnamefont {L.~E.}\ \bibnamefont {Anderson}}, \bibinfo {author}
  {\bibfnamefont {Y.~J.}\ \bibnamefont {Shin}}, \bibinfo {author}
  {\bibfnamefont {S.~Y.}\ \bibnamefont {Lee}}, \bibinfo {author} {\bibfnamefont
  {Y.~H.}\ \bibnamefont {Lee}}, \bibinfo {author} {\bibfnamefont
  {B.}~\bibnamefont {Johnson}}, \bibinfo {author} {\bibfnamefont
  {K.}~\bibnamefont {Watanabe}}, \bibinfo {author} {\bibfnamefont
  {T.}~\bibnamefont {Taniguchi}}, \bibinfo {author} {\bibfnamefont
  {A.}~\bibnamefont {Yacoby}}, \ and\ \bibinfo {author} {\bibfnamefont
  {P.}~\bibnamefont {Kim}},\ }\href {\doibase 10.1038/s41565-021-00861-z}
  {\bibfield  {journal} {\bibinfo  {journal} {Nature Nanotechnology}\ }\textbf
  {\bibinfo {volume} {16}},\ \bibinfo {pages} {563–569} (\bibinfo {year}
  {2021})}\BibitemShut {NoStop}%
\bibitem [{\citenamefont {Nakamura}\ \emph {et~al.}(2023)\citenamefont
  {Nakamura}, \citenamefont {Liang}, \citenamefont {Gardner},\ and\
  \citenamefont {Manfra}}]{nak23}%
  \BibitemOpen
  \bibfield  {author} {\bibinfo {author} {\bibfnamefont {J.}~\bibnamefont
  {Nakamura}}, \bibinfo {author} {\bibfnamefont {S.}~\bibnamefont {Liang}},
  \bibinfo {author} {\bibfnamefont {G.~C.}\ \bibnamefont {Gardner}}, \ and\
  \bibinfo {author} {\bibfnamefont {M.~J.}\ \bibnamefont {Manfra}},\ }\href
  {\doibase 10.1103/PhysRevX.13.041012} {\bibfield  {journal} {\bibinfo
  {journal} {Phys. Rev. X}\ }\textbf {\bibinfo {volume} {13}},\ \bibinfo
  {pages} {041012} (\bibinfo {year} {2023})}\BibitemShut {NoStop}%
\bibitem [{\citenamefont {Appugliese}\ \emph {et~al.}(2022)\citenamefont
  {Appugliese}, \citenamefont {Enkner}, \citenamefont {Paravicini-Bagliani},
  \citenamefont {Beck}, \citenamefont {Reichl}, \citenamefont {Wegscheider},
  \citenamefont {Scalari}, \citenamefont {Ciuti},\ and\ \citenamefont
  {Faist}}]{doi:10.1126/science.abl5818}%
  \BibitemOpen
  \bibfield  {author} {\bibinfo {author} {\bibfnamefont {F.}~\bibnamefont
  {Appugliese}}, \bibinfo {author} {\bibfnamefont {J.}~\bibnamefont {Enkner}},
  \bibinfo {author} {\bibfnamefont {G.~L.}\ \bibnamefont
  {Paravicini-Bagliani}}, \bibinfo {author} {\bibfnamefont {M.}~\bibnamefont
  {Beck}}, \bibinfo {author} {\bibfnamefont {C.}~\bibnamefont {Reichl}},
  \bibinfo {author} {\bibfnamefont {W.}~\bibnamefont {Wegscheider}}, \bibinfo
  {author} {\bibfnamefont {G.}~\bibnamefont {Scalari}}, \bibinfo {author}
  {\bibfnamefont {C.}~\bibnamefont {Ciuti}}, \ and\ \bibinfo {author}
  {\bibfnamefont {J.}~\bibnamefont {Faist}},\ }\href {\doibase
  10.1126/science.abl5818} {\bibfield  {journal} {\bibinfo  {journal}
  {Science}\ }\textbf {\bibinfo {volume} {375}},\ \bibinfo {pages} {1030}
  (\bibinfo {year} {2022})}\BibitemShut {NoStop}%
\bibitem [{\citenamefont {Chklovskii}\ \emph {et~al.}(1992)\citenamefont
  {Chklovskii}, \citenamefont {Shklovskii},\ and\ \citenamefont
  {Glazman}}]{PhysRevB.46.4026}%
  \BibitemOpen
  \bibfield  {author} {\bibinfo {author} {\bibfnamefont {D.~B.}\ \bibnamefont
  {Chklovskii}}, \bibinfo {author} {\bibfnamefont {B.~I.}\ \bibnamefont
  {Shklovskii}}, \ and\ \bibinfo {author} {\bibfnamefont {L.~I.}\ \bibnamefont
  {Glazman}},\ }\href {\doibase 10.1103/PhysRevB.46.4026} {\bibfield  {journal}
  {\bibinfo  {journal} {Phys. Rev. B}\ }\textbf {\bibinfo {volume} {46}},\
  \bibinfo {pages} {4026} (\bibinfo {year} {1992})}\BibitemShut {NoStop}%
\bibitem [{\citenamefont {Mei}\ and\ \citenamefont {Lee}(1983)}]{Mei_1983}%
  \BibitemOpen
  \bibfield  {author} {\bibinfo {author} {\bibfnamefont {W.~N.}\ \bibnamefont
  {Mei}}\ and\ \bibinfo {author} {\bibfnamefont {Y.~C.}\ \bibnamefont {Lee}},\
  }\href {\doibase 10.1088/0305-4470/16/8/010} {\bibfield  {journal} {\bibinfo
  {journal} {Journal of Physics A: Mathematical and General}\ }\textbf
  {\bibinfo {volume} {16}},\ \bibinfo {pages} {1623} (\bibinfo {year}
  {1983})}\BibitemShut {NoStop}%
\bibitem [{\citenamefont {Landau}\ and\ \citenamefont
  {Lifshitz}(1981)}]{Landau1981}%
  \BibitemOpen
  \bibfield  {author} {\bibinfo {author} {\bibfnamefont {L.~D.}\ \bibnamefont
  {Landau}}\ and\ \bibinfo {author} {\bibfnamefont {E.~M.}\ \bibnamefont
  {Lifshitz}},\ }\href {https://books.google.com/books?id=SvdoN3k8EysC} {\emph
  {\bibinfo {title} {Quantum Mechanics: Non-Relativistic Theory}}}\ (\bibinfo
  {publisher} {Elsevier Science},\ \bibinfo {year} {1981})\BibitemShut
  {NoStop}%
\bibitem [{\citenamefont {Tokman}\ \emph
  {et~al.}(2023{\natexlab{a}})\citenamefont {Tokman}, \citenamefont {Behne},
  \citenamefont {Torres}, \citenamefont {Erukhimova}, \citenamefont {Wang},\
  and\ \citenamefont {Belyanin}}]{PhysRevA.107.013721}%
  \BibitemOpen
  \bibfield  {author} {\bibinfo {author} {\bibfnamefont {M.}~\bibnamefont
  {Tokman}}, \bibinfo {author} {\bibfnamefont {A.}~\bibnamefont {Behne}},
  \bibinfo {author} {\bibfnamefont {B.}~\bibnamefont {Torres}}, \bibinfo
  {author} {\bibfnamefont {M.}~\bibnamefont {Erukhimova}}, \bibinfo {author}
  {\bibfnamefont {Y.}~\bibnamefont {Wang}}, \ and\ \bibinfo {author}
  {\bibfnamefont {A.}~\bibnamefont {Belyanin}},\ }\href {\doibase
  10.1103/PhysRevA.107.013721} {\bibfield  {journal} {\bibinfo  {journal}
  {Phys. Rev. A}\ }\textbf {\bibinfo {volume} {107}},\ \bibinfo {pages}
  {013721} (\bibinfo {year} {2023}{\natexlab{a}})}\BibitemShut {NoStop}%
\bibitem [{\citenamefont {Tokman}\ \emph
  {et~al.}(2023{\natexlab{b}})\citenamefont {Tokman}, \citenamefont {Verma},
  \citenamefont {Bohreer},\ and\ \citenamefont
  {Belyanin}}]{PhysRevLett.131.233802}%
  \BibitemOpen
  \bibfield  {author} {\bibinfo {author} {\bibfnamefont {M.}~\bibnamefont
  {Tokman}}, \bibinfo {author} {\bibfnamefont {J.~K.}\ \bibnamefont {Verma}},
  \bibinfo {author} {\bibfnamefont {J.}~\bibnamefont {Bohreer}}, \ and\
  \bibinfo {author} {\bibfnamefont {A.}~\bibnamefont {Belyanin}},\ }\href
  {\doibase 10.1103/PhysRevLett.131.233802} {\bibfield  {journal} {\bibinfo
  {journal} {Phys. Rev. Lett.}\ }\textbf {\bibinfo {volume} {131}},\ \bibinfo
  {pages} {233802} (\bibinfo {year} {2023}{\natexlab{b}})}\BibitemShut
  {NoStop}%
\bibitem [{\citenamefont {Lo}\ \emph {et~al.}(1991)\citenamefont {Lo},
  \citenamefont {Mitchel}, \citenamefont {Perrin}, \citenamefont {Messham},\
  and\ \citenamefont {Yen}}]{PhysRevB.43.11787}%
  \BibitemOpen
  \bibfield  {author} {\bibinfo {author} {\bibfnamefont {I.}~\bibnamefont
  {Lo}}, \bibinfo {author} {\bibfnamefont {W.~C.}\ \bibnamefont {Mitchel}},
  \bibinfo {author} {\bibfnamefont {R.~E.}\ \bibnamefont {Perrin}}, \bibinfo
  {author} {\bibfnamefont {R.~L.}\ \bibnamefont {Messham}}, \ and\ \bibinfo
  {author} {\bibfnamefont {M.~Y.}\ \bibnamefont {Yen}},\ }\href {\doibase
  10.1103/PhysRevB.43.11787} {\bibfield  {journal} {\bibinfo  {journal} {Phys.
  Rev. B}\ }\textbf {\bibinfo {volume} {43}},\ \bibinfo {pages} {11787}
  (\bibinfo {year} {1991})}\BibitemShut {NoStop}%
\end{thebibliography}%
\end{document}